\immediate\write18{makeindex \jobname.nlo -s nomencl.ist -o \jobname.nls}

\RequirePackage{tikz}
\documentclass[sn-mathphys]{sn-jnl}

\UseRawInputEncoding
\usepackage{amsmath}
\usepackage{graphicx}
\usepackage{subfigure}
\usepackage{enumerate}
\usepackage{float}
\usepackage[percent]{overpic}
\usepackage{sidecap}
\usepackage{tipa}
\usepackage{epstopdf}
\usepackage{booktabs}
\usepackage{hyperref}
\usepackage{lineno}
\usepackage{appendix}
\usepackage{arydshln}
\usepackage{mathtools}
\usepackage{siunitx}

\usepackage{scalerel,stackengine,amsmath}
\newcommand\equalhat{\mathrel{\stackon[1.5pt]{=}{\stretchto{%
    \scalerel*[\widthof{=}]{\wedge}{\rule{1ex}{3ex}}}{0.5ex}}}}

\usepackage{tikz}
\DeclareRobustCommand*\circled[1]{\tikz[baseline=(char.base)]{
\node[shape=circle,draw,inner sep=1pt] (char) {#1};}}

\DeclareMathOperator{\sign}{sign}
\DeclareSIUnit{\clicks}{Clicks}

\usepackage{framed} 
\usepackage[nomentbl,stdsubgroups,nocfg]{nomencl} 

\makenomenclature

\jyear{2021}%

\theoremstyle{thmstyleone}%
\theoremstyle{thmstyletwo}%

\theoremstyle{thmstylethree}%

\begin{document}

\title[A universal nonlinear model for the dynamic behaviour of shock absorbers]{A universal nonlinear model for the dynamic behaviour of shock absorbers}


\author*[1,2]{\fnm{Lukas} \sur{Schickhofer}}\email{lukas.schickhofer@ktm.com; lukas.schickhofer13@alumni.imperial.ac.uk}



\affil*[1]{\orgname{KTM TECHNOLOGIES}, \orgaddress{\city{Salzburg}, \postcode{AT-5081}, \country{Austria}}}

\affil*[2]{\orgname{KTM Research \& Development}, \orgaddress{\city{Mattighofen}, \postcode{AT-5230}, \country{Austria}}}



\abstract{Modern hydraulic shock absorbers display a wealth of nonlinear effects such as hysteresis and instabilities at high flow rates.
Despite their wide application in practically all vehicles, both on- and off-road, a universal analytical model that captures the essential shock absorber dynamics and compressibility effects for various common damper architectures is lacking.
This paper presents such a model and derives its system of equations from first principles for dampers of monotube- and piggyback-type.
By applying the model to a typical suspension configuration, all relevant system variables, such as pressure drop, shim stack deflection, and damping force, are computed.
Nonlinear oscillations and hysteresis loops, which might prove dangerous during operation, can be predicted effortlessly.
The results achieved with the mathematical model are validated and agree well with test bench measurements. Furthermore, the presented approach is shown to be easily modifiable to describe the physical processes within other hydraulic shock absorber geometries in use today.}




\maketitle


\newpage

\nomenclature[aA]{$A$}{Chamber cross-sectional area}{\square\meter}{}
\nomenclature[aAb]{$A_{b}$}{Bleed area}{\square\meter}{}
\nomenclature[aAp]{$A_{p}$}{Area of acting pressure}{\square\meter}{}
\nomenclature[aAv]{$A_{v}$}{Section of valve opening}{\square\meter}{}
\nomenclature[aQb]{$Q_{b}$}{Bleed flow rate}{\cubic\meter\per\second}{}
\nomenclature[aQl]{$Q_{l}$}{Leakage flow rate}{\cubic\meter\per\second}{}
\nomenclature[aQv]{$Q_{v}$}{Valve flow rate}{\cubic\meter\per\second}{}
\nomenclature[aL]{$L$}{Chamber length}{\meter}{}
\nomenclature[ab]{$b$}{Leakage gap width}{\meter}{}
\nomenclature[al]{$l$}{Leakage gap length}{\meter}{}
\nomenclature[ac]{$a_{c}$}{Clamp radius}{\meter}{}
\nomenclature[a]{$a$}{Shim radius}{\meter}{}
\nomenclature[at]{$t$}{Shim thickness}{\meter}{}
\nomenclature[am]{$m$}{Mass}{\kilogram}{}
\nomenclature[ak]{$k$}{Stiffness}{\newton\per\meter}{}
\nomenclature[ac]{$c$}{Damping coefficient}{\newton\second\per\meter}{}
\nomenclature[aD]{$D$}{Flexural rigidity}{\newton\square\meter}{}
\nomenclature[aE]{$E$}{Young's modulus}{\pascal}{}
\nomenclature[aCd]{$C_{d}$}{Discharge coefficient}{-}{}
\nomenclature[aCf]{$C_{f}$}{Moment coefficient}{-}{}
\nomenclature[aq]{$q$}{Load per unit area}{\pascal}{}
\nomenclature[aw]{$w$}{Unit line load}{\newton\per\meter}{}
\nomenclature[aF0]{$F_{0}$}{Pretension force}{\newton}{}
\nomenclature[aFp]{$F_{p}$}{Pressure force}{\newton}{}
\nomenclature[aFm]{$F_{m}$}{Momentum force}{\newton}{}
\nomenclature[aFi]{$F_{i}$}{Impact force}{\newton}{}
\nomenclature[ax]{$x$}{Displacement}{\meter}{}
\nomenclature[ap]{$p$}{Pressure}{\pascal}{}

\nomenclature[ga]{$\alpha$}{Flow area proportionality factor}{-}{}
\nomenclature[gaT]{$\alpha_{T}$}{Thermal expansion coefficient}{-}{}
\nomenclature[gr]{$\rho$}{Density}{\kilogram\per\cubic\meter}{}
\nomenclature[gm]{$\mu$}{Dynamic viscosity}{\kilogram\per(\meter\second)}{}
\nomenclature[gc]{$\beta$}{Compressibility}{\square\meter\per\newton}{}
\nomenclature[gd]{$\delta$}{Shim deflection}{\meter}{}
\nomenclature[gp]{$\nu$}{Poisson's ratio}{-}{}

\nomenclature[xt]{$\sim$}{Check valve piston quantity}{}{}

\nomenclature[zc]{$c$}{Compression quantity}{}{}
\nomenclature[zd]{$d$}{Damper compartment quantity}{}{}
\nomenclature[zr]{$r$}{Rebound quantity}{}{}
\nomenclature[zg]{$g$}{Gas reservoir quantity}{}{}
\nomenclature[z0]{$0$}{Reference value}{}{}

{
\let\cleardoublepage\clearpage
\begin{framed}
\printnomenclature
\end{framed}
}

\vspace{0.5cm}


\section{Introduction}
\label{Sec:1a}

In order to protect drivers and passengers from the impact of road irregularities, an effective suspension is a vital part of practically all vehicles. Thereby, it does not only guarantee comfort during driving across uneven ground, but also ensures safety in potentially dangerous situations, where wheel grip and road contact might be compromised.
Thus, suspension design and modelling has become increasingly relevant for a wide range of applications, from race cars, motorcycles, and bikes, to trains \citep{dixon2008shock,boggs2010efficient,wang2017rail}.
The core components of modern suspension systems are hydraulic shock absorbers, also known as dampers, which provide a variable damping force due to displacement of fluid through orifices of changing cross sections. These are further supported by simple elastic springs of constant stiffness that offer linear forces against the displacement of moving parts and axles. While springs are easy to model, they fail to adjust to different stroking amplitudes and frequencies, and would not adapt to a variation in driving scenarios. This problem is crucially solved by the hydraulic shock absorber, which allows for several ways to control the damping characteristics by controlled and variable pressure losses across its compartments. However, this freedom in designing an ideal configuration comes at the cost of increased complexity and nonlinearity in its function. 
Therefore, efforts have been made to derive mathematical models that reliably predict the behaviour of a shock absorber under realistic conditions.
This is particularly important for high-performance applications, such as in motorsports, where quick adjustments need to be made under time constraints and without extensive testing, but also highly relevant for the first phase of development of general-purpose vehicles for serial production.
However, as stated by various authors, research into the dependencies of crucial shock absorber characteristics, such as damping forces, pressure drops, or shim stack stiffness, on the large set of modifiable parameters of the system is limited \citep{dixon2008shock,farjoud2012nonlinear,skavckauskas2017development}.

Lang \cite{lang1977study} was the first to derive a set of suitable equations for the modelling of the unsteady valve flow within an automotive shock absorber. He found a way to connect the volumetric flow rates and pressure differences across the damper volumes and used empirically established values for the associated discharge coefficients. These models were capable of predicting the forces acting on the shock absorber piston and subsequently the damping performance, even at higher stroking frequencies.
Later on, Reybrouck \cite{reybrouck1994non} formulated a nonlinear model of a shock absorber with pressure drops depending on the fixed and variable constrictions of the piston valves. The specific system parameters were defined based on test bench measurements and experiments were used to validate the mathematical model. The model was based on semi-empirical coefficients and thus did not allow investigation of the effects of internal modifications on shock absorber performance. 
Similarly, Lee \cite{lee1997numerical} modelled a monotube damper by a low-order computational model, which considered spring-loaded disk valves, but did not include the effect of shim stacks.
Another approach was taken by Talbott and Starkey \cite{talbott2002experimentally}, who devised a nonlinear algebraic system of equations from physical principles and reached good agreement with measurements. Using this model, they were able to perform parametric studies and came to conclusions about the inner functional dependencies of a monotube damper, for instance that its variation in rebound pressure dominates the overall pressure drop.

The underlying issue of the models mentioned above was their common assumption of the incompressibility of the mineral oil in the shock absorber chambers. This assumption allows to enforce the conservation of volume of the fluid and further leads to the formulation of a system of algebraic equations. Such a system can then be solved with suitable methods, such as the Newton-Raphson method (as demonstrated e.g. by Rhoades \cite{rhoades2006development} for the improved model of Talbott and Starkey \cite{talbott2002experimentally}), to find the system's variables during each piston displacement in time.
However, important compressibility effects, such as hysteresis, where the damping force varies between compression and rebound stage, are neglected. This problem was consequently taken up by Duym et al. \cite{duym1997evaluation,duym1997physical}, who evaluated previous algebraic approaches and suggested to model hysteresis by including two dominant effects: The oil compressibility and the compressibility of a variable gas phase present as bubbles from the adjacent nitrogen reservoir chamber. The latter is a particular issue of dual-tube shock absorbers and has been accounted for by using Henry's law for the gas solubility and subsequent isothermal expansion and compression.

Another crucial extension to previous modelling approaches was the correct calculation of the stiffness of the shim stack or valves that are used for the dynamic flow restriction. Recently, several validated physical models made use of linearization and treated the shim stack as a system exposed to static loads computed by the force method \citep{skavckauskas2017development,xu2018hybrid}. In these works, singular loads are superimposed to give the total shim stack deflection, but a rigorous methodology for its numerical implementation is lacking. Czop et al. \cite{czop2009simplified} evaluated simplified, reduced-order models of shim stacks and compared linear with more advanced nonlinear approaches. They analysed the relative strengths and deviations of each method at high pressures, but did not provide a validation or comparison to experimental data. Farjoud et al. \cite{farjoud2012nonlinear} on the other hand used the Rayleigh-Ritz approximation for the minimization of the total potential energy of a shim stack in order to find its state of deflection at given time-dependent boundary conditions. However, the large number of unknown coefficients of the method, which need to be calculated by a nonlinear optimization algorithm, makes it impractical for quick computations.

Additionally, simulations using the finite-element method (FEM), as well as coupled numerical solvers for fluid-structure interaction (FSI) opened up new possibilities for the computation of shim stack and valve deflection in realistic three-dimensional geometries \citep{bell2017numerical,xu2018hybrid,hofmann2018investigation,schickhofer2022fluid}. 
Such studies could predict the exact flow losses and areas of acting pressure within the shock absorber piston and shim stack -- parameters that are typically not captured in reduced-order models.
Nevertheless, despite the large increase of computing power in recent years, high-order simulation methods are often impractical for the quick and reliable adjustment of shock absorbers in pre-testing or race track applications, especially due to the vast amount of possible parameter combinations. Thus, there is still a large demand for mathematical models based on fundamental physical principles that are suitable for fast numerical integration and that allow for important conclusions regarding the functional dependencies and sensitivities to parameter changes of a shock absorber.


\section{Motivation}
\label{Sec:1b}

The motivation for this work is two-fold:
\begin{enumerate}[(i)]
\item The formulation of a system of nonlinear coupled differential equations to capture all important dynamic effects during shock absorber operation and to allow for a quick computation of all system variables.
\item The definition of a mathematical modelling approach that can be applied for a wide range of shock absorber types.
\end{enumerate}
The first point comes from the fact that crucial nonlinear effects, such as hysteresis, or the occurrence of instabilities are not captured with an algebraic model. However, the majority of shock absorber models, as outlined in Sec. \ref{Sec:1a} are algebraic in nature, or are not given in the form of a closed, first-order system, which can be used for future studies of its dynamics.
The second point derives from the variety of shock absorber architectures currently in use. It is therefore a dedicated goal of this work to establish a modelling approach, which can be easily adapted to the analysed system.


\section{Methodology}
\label{Sec:2}

In the following, the basic setup of shock absorbers is explained in more detail (cf. Sec. \ref{Sec:2A}), alongside the essential definitions of its physical quantities with the applied strategy for its modelling (cf. Sec. \ref{Sec:2B}). Finally, the complete systems of equations for the considered shock absorber types are presented in Sec. \ref{Sec:2C}-\ref{Sec:2D}.

\subsection{General setup of shock absorbers}
\label{Sec:2A}

\begin{figure*}[htbp]
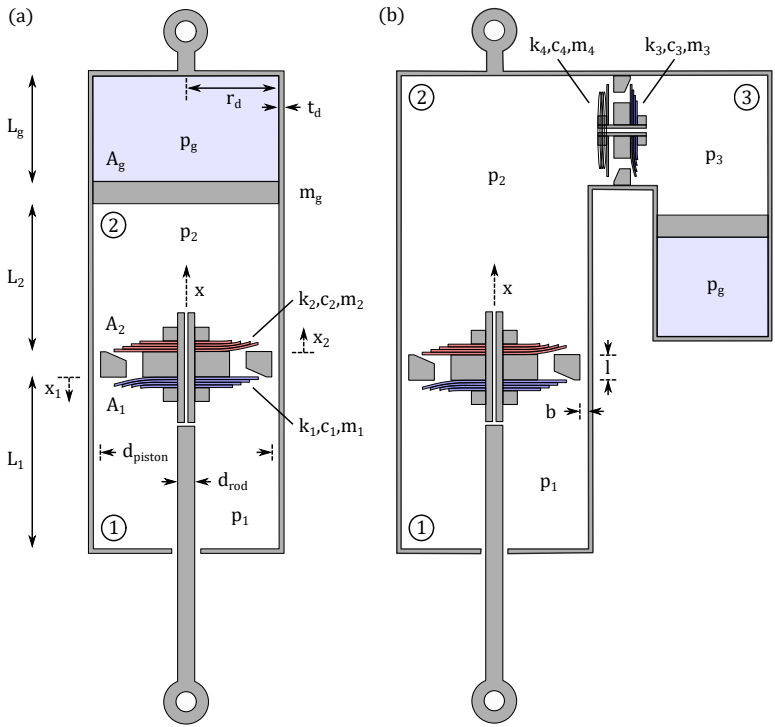

\begin{centering}
\begin{overpic}[width=0.85\textwidth]{Figure1.pdf}
\end{overpic}
\par
\end{centering}
\caption{Sketch of a monotube version (a) and a piggyback version (b) of a hydraulic shock absorber. The relevant quantities for the model formulation are indicated together with the compression chamber \circled{1}, the rebound chamber \circled{2}, the second compression chamber \circled{3} in case of the piggyback geometry, and the gas reservoir.
The shims or valves engaged during compression are coloured blue, while the ones engaged during rebound are coloured red.}
\label{Fig:2A-1} 
\end{figure*}

Shock absorbers, or automotive hydraulic dampers, come in various forms and shapes. However, the most common type used today is the \emph{monotube damper}, as shown in Fig. \ref{Fig:2A-1}(a). In its most general setup, it consists of a metal cylinder (typically of stainless steel) that contains an axially moving piston surrounded by damping fluid such as mineral oil. The piston is connected to a rod that takes up displacements (e.g. due to road irregularities or impact) and also separates the inside of the cylinder into a compression and rebound chamber. Since the compressibility of the oil is small, an additional gas reservoir filled with gas of inert-like characteristics (e.g. nitrogen) is necessary in order to accommodate the rapid change in volume during operation.
Additionally, the piston is equipped on both sides with a stack of metal shims, which are typically made of a steel alloy of high yield strength. The damping fluid flows through valve ports of different cross-sectional areas and acts against the resistance of the shim stacks. Depending on the number of shims, as well as on the size of a small constant bleed orifice, the pressure drop across the piston during the compression and rebound stage can be adjusted.

Another widely applied shock absorber type, particularly for rear-suspension applications of motorcycles, is the \emph{piggyback damper} (cf. Fig. \ref{Fig:2A-1}(b)).
The name derives from the additional compartment for the nitrogen chamber, which is separated from the main cylinder by a much smaller check valve piston that controls the flow direction. It is typically equipped with a shim stack in compression direction and a simple check valve loaded by a spring of small stiffness, which ensures quick response to changes in the flow direction and little delay during loading of the compression stacks.

\begin{figure*}[htbp]
\begin{centering}
\begin{overpic}[width=1.00\textwidth]{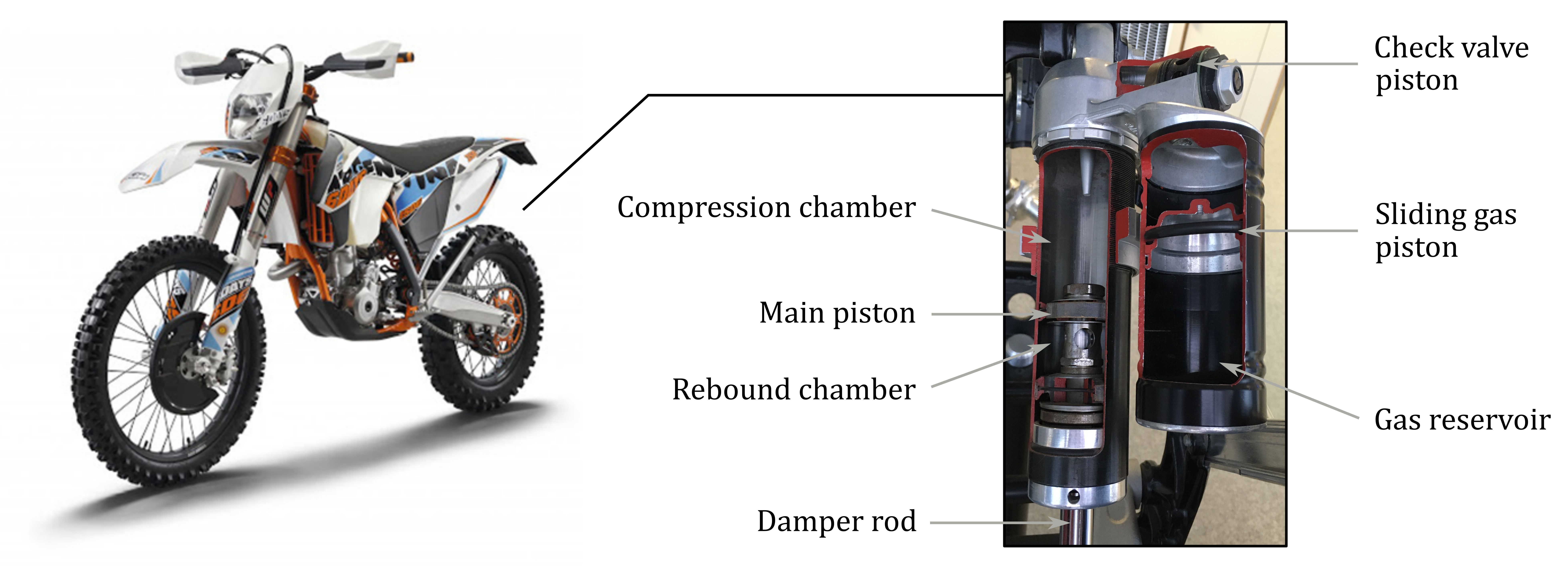}
\end{overpic}
\par
\end{centering}
\caption{Cross section of a piggyback damper as applied within an off-road motorcycle back suspension system. The components as introduced in Fig. \ref{Fig:2A-1} are indicated. (Photos courtesy of \emph{WP Suspension} and \emph{KTM}.)}
\label{Fig:2A-1a} 
\end{figure*}

The piggyback damper has several crucial advantages and differences in the functional dependencies of its parameters when compared to the classic monotube damper, which are discussed in Sec. \ref{Sec:4}-\ref{Sec:5}. It is therefore applied in various vehicles exposed to high-impact scenarios. An example is the back suspension system of modern motorcycles, as pictured in Fig. \ref{Fig:2A-1a}.
Despite their apparent differences, both the monotube and the piggyback architecture of shock absorbers can be analytically described with the proposed model, which is presented in its specific formulations below.

\begin{figure*}[htbp]
\begin{centering}
\begin{overpic}[width=0.85\textwidth]{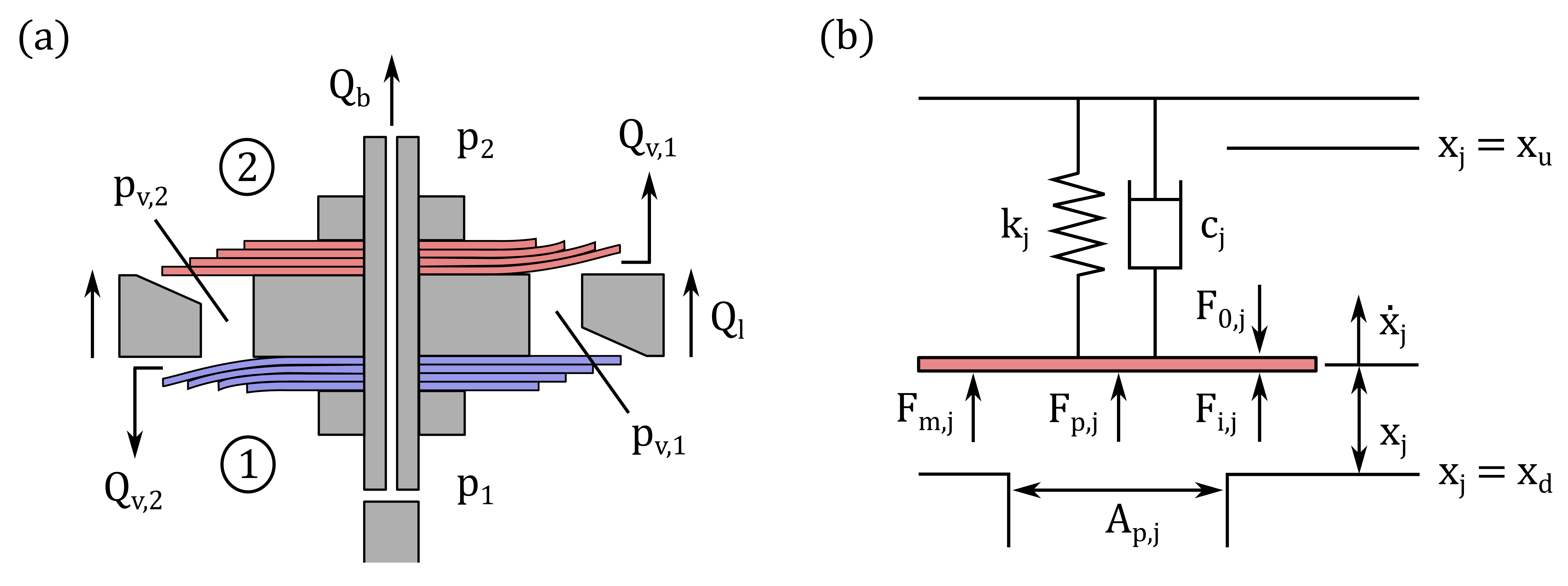}
\end{overpic}
\par
\end{centering}
\caption{Diagram of volumetric flow rates across the main piston (a). The actual flow direction changes due to the pressure difference across the piston, which is defined in Sec. \ref{Sec:2B-1}. During opening, various forces are acting on the valves or shims, which are characterised by a certain stiffness $k_{j}$ and damping coefficient $c_{j}$ (b). The discontinuities at the lower bound of the domain at $x_{j}=x_{d}$ and at the upper bound at $x_{j}=x_{u}$ are indicated. At these positions the motion of the driven oscillator is counteracted by the applied impact force of Eq. (\ref{Eq:2B-2-3}).}
\label{Fig:2A-2} 
\end{figure*}

\subsection{Governing equations}
\label{Sec:2B}

In order to obtain a system of equations that describes all relevant quantities of the shock absorber dynamics for both the monotube and the piggyback type, several basic variables and their connection to each other need to be defined. This is presented in the following sections.

\subsubsection{Conservation laws}
\label{Sec:2B-1}

First, all volumetric flows in and out of the considered domain need to be defined to obtain a conservation equation for each control volume. These are essentially the chambers \circled{1}, \circled{2}, \circled{3}, and the gas reservoir in Fig. \ref{Fig:2A-1}.
Starting from the general form of the Bernoulli equation
\begin{equation}
p + \frac{\rho u^2}{2} + \rho g h = const.  \nonumber
\label{Eq:2B-1-1}
\end{equation}
and by neglecting pressure differences induced by gravity, the volumetric flow rate between two points of reference 1 and 2 is found as
\begin{equation}
Q\left(p_{1},p_{2}\right) = A \cdot \sign\left(p_{1}-p_{2}\right) \sqrt{\frac{2\left|p_{1}-p_{2}\right|}{\rho}},  \nonumber
\label{Eq:2B-1-2}
\end{equation}
where the cross-sectional area $A$ is considered constant.
Steady and incompressible flow is assumed in the small flow domain between the adjacent compression and rebound chambers of each piston, such that the application of the Bernoulli equation is indeed valid in this context. Moreover, potential transition to turbulent flow in the connecting piston orifices is modelled by the adapted discharge coefficients defined by Eq. (\ref{Eq:2B-1-5})-(\ref{Eq:2B-1-6}) further below.

Applied to the flow within a shock absorber, such as the monotube damper in Fig. \ref{Fig:2A-1}(a), one can identify flow rates through the valve openings
\begin{align}
Q_{v,1}\left(x_{2},p_{2},p_{v,1}\right) &= \underbrace{\alpha \pi d_{max,2} x_{2}}_{A_{v,2}\left(x_{2}\right)} C_{d,v,1} \cdot \sign\left(p_{2}-p_{v,1}\right) \sqrt{\frac{2\left|p_{2}-p_{v,1}\right|}{\rho}}, \label{Eq:2B-1-3a} \\
Q_{v,2}\left(x_{1},p_{v,2},p_{1}\right) &= \underbrace{\alpha \pi d_{max,1} x_{1}}_{A_{v,1}\left(x_{1}\right)} C_{d,v,2} \cdot \sign\left(p_{v,2}-p_{1}\right) \sqrt{\frac{2\left|p_{v,2}-p_{1}\right|}{\rho}}, \label{Eq:2B-1-3b}
\end{align}
where the discharge coefficient $C_{d,v,j}$ is due to the \emph{vena contracta effect}, which leads to the contraction of a flow from its initial diameter to a smaller one when exiting or entering an orifice.
It is important to note that the cross-sectional area of the ejected flow $A_{v,j}\left(x_{j}\right)$ is now dependent on the axial opening during compression or rebound due to deflection of the considered shim stack or check valve $j$ ($j=1,2$).
Moreover, the proportionality factor $\alpha$ takes into account that the actual flow area is only a fraction of the circumferential surface $\pi d_{max,j} x_{j}$, where $d_{max,j}$ is the diameter of the first shim of stack $j$ adjacent to the piston.
The factor $\alpha$ is essentially defined by the port geometry of the piston, i.e. what fraction of the piston surface is taken up by the ports.
Since the fluid already undergoes a pressure drop when entering the valve ports, this needs to be accounted for via a decreased valve pressure $p_{v,j}$ as follows:
When the oil enters the port for valve $j$, the related volumetric flow rate is the same as the one exiting over the valve on the other side of the piston. Thus, an alternative expression can be found as
\begin{equation}
Q_{v,j} \simeq Q_{c,j}\left(p_{j},p_{v,j}\right)  = A_{c,j}C_{d,c,j} \cdot \sign\left(p_{j}-p_{v,j}\right) \sqrt{\frac{2\left|p_{j}-p_{v,j}\right|}{\rho}},
\label{Eq:2B-1-3c}
\end{equation}
with the port channel cross-sectional area $A_{c,j}$ and the port discharge coefficient $C_{d,c,j}$.
Using Eq. (\ref{Eq:2B-1-3a})-(\ref{Eq:2B-1-3c}), the valve port pressures $p_{v,j}$ can be calculated as
\begin{align}
p_{v,1} &= \frac{C_{d,c,1}^{2}A_{c,1}^{2}p_{1} + C_{d,v,1}^{2}A_{v,1}^{2}p_{2}}{C_{d,c,1}^{2}A_{c,1}^{2} + C_{d,v,1}^{2}A_{v,1}^{2}},  \label{Eq:2B-1-3d}  \\
p_{v,2} &= \frac{C_{d,c,2}^{2}A_{c,2}^{2}p_{2} + C_{d,v,2}^{2}A_{v,2}^{2}p_{1}}{C_{d,c,2}^{2}A_{c,2}^{2} + C_{d,v,2}^{2}A_{v,2}^{2}}.  \label{Eq:2B-1-3e}
\end{align}
In most applications the channel area is equal to the area of acting pressure on the adjacent shim stack (e.g. $A_{c,1}=A_{p,2}$). However, this is not always the case and in several modern shock absorbers the seat area of the shim stack is enlarged to allow for targeted modification of the high-speed damping behaviour.
The valve flow rates are also depicted in Fig. \ref{Fig:2A-2}(a), with the flow direction crucially defined by the sign of the pressure difference, as given by Eq. (\ref{Eq:2B-1-3a})-(\ref{Eq:2B-1-3b}).

Furthermore, there is a bleed flow rate through a small constant orifice in the centre of the piston
\begin{equation}
Q_{b}\left(p_{2},p_{1}\right) = A_{b} C_{d,b} \cdot \sign \left(p_{2}-p_{1}\right) \sqrt{\frac{2\left|p_{2}-p_{1}\right|}{\rho}},
\label{Eq:2B-1-4}
\end{equation}
where the bleed area $A_{b}$ can be adjusted by a needle to control the damping behaviour particularly at low damper excitation and small total flow rates.

\begin{figure*}[htbp]
\begin{centering}
\begin{overpic}[width=0.80\textwidth]{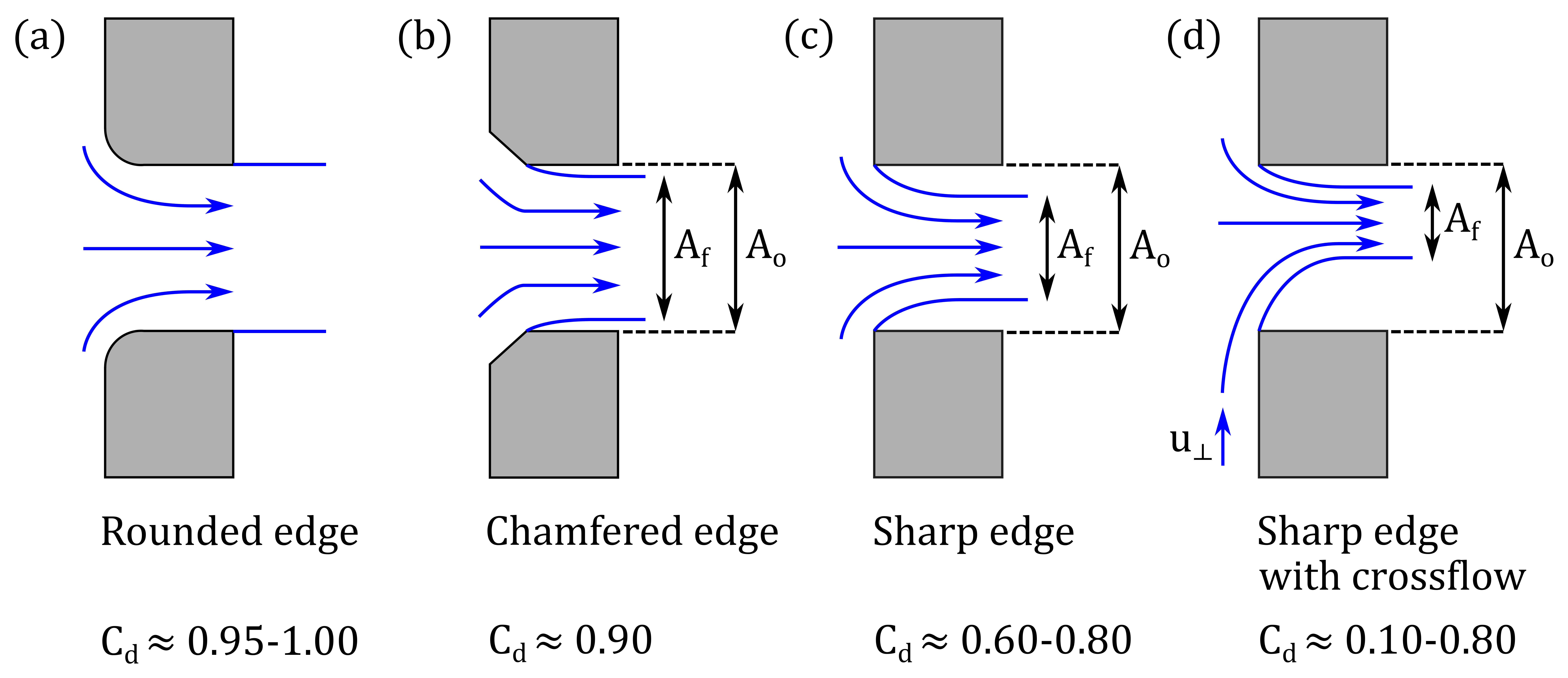}
\end{overpic}
\par
\end{centering}
\caption{Discharge coefficients $C_{d}$ as the ratio of actual flow section $A_{f}$ over orifice section $A_{o}$ for various edge geometries. There is a wide range of values from rounded (a) to chamfered (b) and sharp edges (c), which are based on experimental measurements \citep{dixon2008shock,feseker2018experimental}. In case of an additional shear flow component $u_{\perp}$, the discharge coefficient can drop significantly (d).}
\label{Fig:2B-1-1} 
\end{figure*}

The discharge coefficient varies depending on the flow being laminar or turbulent and is usually determined by experiments. 
In order to be able to apply Bernoulli's equation to the unsteady flow within a shock absorber and to take into account transition to turbulence, a continuously differentiable, analytic expression for the discharge coefficient is applied with
\begin{equation}
\begin{split}
C_{d,v} &= C_{d,v,max}\cdot\tanh\left(\frac{2\operatorname{Re}}{\operatorname{Re}_{c,v}}\right)=C_{d,v,max}\cdot\tanh\left(\frac{2d_{v}}{\nu \operatorname{Re}_{c,v}}\sqrt{\frac{2\left|\Delta p\right|}{\rho}}\right) \\
              &= C_{d,max,v}\cdot\tanh\left(\frac{4x}{\nu \operatorname{Re}_{c,v}}\sqrt{\frac{2\left|\Delta p\right|}{\rho}}\right).
\label{Eq:2B-1-5} 
\end{split}
\end{equation}
for the discharge coefficient related to variable valve opening $x$ and
\begin{equation}
C_{d,b}=C_{d,b,max}\cdot\tanh\left(\frac{2\operatorname{Re}}{\operatorname{Re}_{c,b}}\right)=C_{d,b,max}\cdot\tanh\left(\frac{2d_{b}}{\nu \operatorname{Re}_{c,b}}\sqrt{\frac{2\left|\Delta p\right|}{\rho}}\right),
\label{Eq:2B-1-6} 
\end{equation}
for the constant bleed orifice discharge coefficient \cite{mccloy1980control}.
The values $\operatorname{Re}_{c,b} \approx 1000$ and $\operatorname{Re}_{c,v} \approx 100$ for the critical Reynolds number indicate the transition from laminar to turbulent flow. For the small gap openings of the valve, the transition occurs typically at much smaller Reynolds number ($\operatorname{Re}_{c,v} \ll \operatorname{Re}_{c,b}$).
Typical empirically determined values for the maximum discharge coefficients $C_{d,max}$ of suspension flows are within the range of $0.6$--$0.7$ \citep{dixon2008shock}. Segel and Lang \cite{segel1981mechanics} for instance proposed $C_{d,max}=0.7$ for regular jets within a shock absorber. However, the maximum discharge coefficients, as demonstrated in Fig. \ref{Fig:2B-1-1}, are crucially dependent on the piston geometry and might be considerably lower (or higher) than the above values, especially for cases of chamfered edges at the entry (or the geometric expansion of the discharge area at the exit of a port channel) \citep{dixon2008shock}. Furthermore, an extensive experimental study by Feseker et al. \cite{feseker2018experimental} on discharge coefficients of circular orifices with additional shear flow  showed a sharp drop of fluid discharge with increasing inlet crossflow velocity. Thus, a definition of the exact maximum discharge coefficients for a given shock absorber geometry can often only be obtained empirically via validation and comparison to test bench measurements, as demonstrated in Sec. \ref{Sec:3}.

In addition to the above flow rates, there is a small leakage flow through the gap between the moving piston and the surrounding cylinder walls, which can be modelled as a flow between two parallel plates, as suggested by Lang \citep{lang1977study} and the analytical solution by Munson, Young, and Okiishi \citep{munsonfundamentals}. 
In the approximation applied here, the flow is considered laminar due to the small clearance and a solution to the Navier-Stokes equations under these boundary conditions is found as
\begin{equation}
Q_{l}\left(p_{2},p_{1}\right) = \left[ \frac{\left(p_{2}-p_{1}\right)b^{3}}{12 \mu l} + \dot{x}\frac{b}{2} \right] \pi d_{piston}.
\label{Eq:2B-1-7}
\end{equation}
Here, the width $b$ and length $l$ define the gap region between the piston and the shock absorber cylinder.
Modern variants of shock absorbers are equipped with rubber o-rings that press the piston outer edge against the cylinder such that the edge is close to zero ($b \approx 0$). 

Finally, the motion of the main piston within the shock absorber causes a certain displacement of fluid, which can be accounted for by the flow rate term $A_{j}\dot{x}$ with the surface area $A_{j}$ depending on compression or rebound side and the rod velocity $\dot{x}$.
For the rebound side this translates to the section $A_{1}=A_{piston}-A_{rod}$,
to account for the diameter of the rod, while for the compression side it is equal to the piston diameter $A_{2}=A_{piston}$.

The above definitions of the volumetric flow rates are now used for the balance equation of the damping oil within the considered control volume.
In case of an incompressible fluid, conservation of fluid volume would be ensured at all time. However, since the mineral oil used as damping fluid in the various chambers is in general slightly compressible, this needs to be considered in the model formulation due to crucial effects of compressibility on shock absorber function, such as hysteresis. This stands in contrast to the initial notion above, where fluid incompressibility is assumed to obtain a relationship between flow rate and pressure drop from the Bernoulli equation. Nonetheless, this simplifying assumption above is only taken in the short connective orifices, while the oil in the chambers \circled{1}, \circled{2}, \circled{3} is subjected to the bulk of the pressure forces.
Thus, and by accounting for compressibility within those control volumes, the continuity equation in integral form
\begin{equation}
\frac{\partial\rho}{\partial t}\int_{V_{j}}dV+\int_{A_{j}}\rho\vec{u}\cdot\vec{n}dA = 0  \nonumber
\label{Eq:2B-1-8} 
\end{equation} 
is applied to the occurring chamber volumes $j$ to give the general expression
\begin{equation}
V_{j}\left(x\right)\frac{\partial\rho}{\partial t} = \rho\left(\sum_{j=1,2} Q_{v,j} + Q_{b} + Q_{l} - A_{j}\dot{x}\right).
\label{Eq:2B-1-9} 
\end{equation}
Since we assume isothermal conditions, the fluid compressibility can be introduced as
 \begin{equation}
\beta=\frac{1}{K} = \frac{1}{\rho}\left[ \frac{d\rho}{dp} \right]_{T=T_{0}},
\label{Eq:2B-1-10} 
\end{equation}
which is the inverse bulk modulus $K$.
This allows to rewrite Eq. (\ref{Eq:2B-1-9}) as a differential equation in terms of the pressure changes within the control volume $V_{j}$.
For the rebound chamber \circled{1} and the compression chamber \circled{2} of the monotube damper in Fig. \ref{Fig:2A-1}(a), one gets
\begin{align}
\dot{p}_{1} &= \frac{1}{\beta \left[A_{1}\left(L_{1}+x\right)\right]} \left( Q_{v,1} + Q_{v,2} + Q_{b} + Q_{l} - A_{1}\dot{x} \right),   \label{Eq:2B-1-11a}  \\
\dot{p}_{2} &= \frac{1}{\beta \left[A_{2}\left(L_{2}-x\right)+A_{g}x_{g}\right]} \left( - Q_{v,1} - Q_{v,2} - Q_{b} - Q_{l} + A_{2}\dot{x} - A_{g}\dot{x}_{g} \right),  \label{Eq:2B-1-11b}
\end{align}
with the definitions for the volumetric flow rates $Q_{v,1}$, $Q_{v,2}$, $Q_{b}$, and $Q_{l}$ as introduced above.

In Eq. (\ref{Eq:2B-1-11b}) the displacement of the floating gas piston $x_{g}$ is taken into account.
The gas compartment, also shown in Fig. \ref{Fig:2A-1} serves the purpose of absorbing the volume displaced by the main piston.
Since the oil is only slightly compressible, the displacement leads mainly to a compression of the contained gas.
For an adiabatic process involving ideal gas, where the transfer of heat with surroundings is negligible, the associated pressure change follows directly from the first law of thermodynamics with only pressure-volume work:
\begin{align}
Vdp &= - pdV,  \nonumber  \\
\frac{dV}{dt}dp + V\frac{dp}{dt} &= - \frac{dp}{dt}dV - p\frac{dV}{dt},  \nonumber  \\
\underbrace{\frac{dp}{dt}}_{\dot{p}_{g}}\underbrace{\left(V+dV\right)}_{V_{g}-A_{g}x_{g}} &= - \underbrace{\frac{dV}{dt}}_{-A_{g}\dot{x}_{g}}\underbrace{\left(p+dp\right)}_{\gamma p_{g}},  \nonumber  \\
\dot{p}_{g} &= - \gamma \frac{p_{g}\left(-A_{g}\dot{x}_{g}\right)}{A_{g}\left(L_{g}-x_{g}\right)}.  \label{Eq:2B-1-12}
\end{align}
Here, $A_{g}$ is the cross-sectional area in the gas reservoir, $L_{g}$ its initial length, and $\gamma$ the adiabatic constant, which is equal to $7/5=1.4$ for diatomic gases such as nitrogen or oxygen.
With Eq. (\ref{Eq:2B-1-11a})-(\ref{Eq:2B-1-12}) the pressure changes within the shock absorber system are sufficiently defined.

\subsubsection{Forces}
\label{Sec:2B-2}

The shim stacks and valves within the shock absorber, which are engaged during the compression and rebound stage, control the flow through the ports and thus the pressure drop across the piston. They are objected to various forces that need to be accounted for and balanced accordingly to find an equation of motion (cf. Fig. \ref{Fig:2A-2}(b)).

First, there is a pressure force by the damper fluid
\begin{align}
F_{p,1}\left(p_{v,2},p_{1}\right) = A_{p,1} \left(p_{v,2}-p_{1}\right),  \label{Eq:2B-2-1a}  \\
F_{p,2}\left(p_{v,1},p_{2}\right) = A_{p,2} \left(p_{v,1}-p_{2}\right),  \label{Eq:2B-2-1b}  
\end{align}
which acts on the first shim over a surface $A_{p,j}$.

Furthermore, there is a momentum force, which follows from the impact of the fluid at the shims and the subsequent redirecting of the jet flow exiting the valve ports. It is found by conservation of momentum as
\begin{align}
F_{m,1}\left(x_{1},p_{v,2},p_{1}\right) &= C_{f} \rho \frac{Q_{v,2}^{2}\left(x_{1},p_{v,2},p_{1}\right)}{A_{p,1}} \nonumber  \\
                                                                  &= \frac{C_{f} \rho}{A_{p,1}} \sign\left(p_{v,2}-p_{1}\right) \left( \alpha \pi d_{max,1} x_{1} C_{d} \cdot \sqrt{\frac{2\left|p_{v,2}-p_{1}\right|}{\rho}} \right)^{2},  \label{Eq:2B-2-2a}  \\
F_{m,2}\left(x_{2},p_{v,1},p_{2}\right) &= C_{f} \rho \frac{Q_{v,1}^{2}\left(x_{2},p_{v,1},p_{2}\right)}{A_{p,2}} \nonumber  \\
                                                          &= \frac{C_{f} \rho}{A_{p,2}} \sign\left(p_{v,1}-p_{2}\right) \left( \alpha \pi d_{max,2} x_{2} C_{d} \cdot \sqrt{\frac{2\left|p_{v,1}-p_{2}\right|}{\rho}} \right)^{2},  \label{Eq:2B-2-2b}  
\end{align}
where $C_{f}$ is a weighting coefficient for the momentum term, since the exact flow conditions within the valve ports are unknown.
Experiments by Lang \citep{lang1977study} on shock absorbers with exceptionally high excitation frequencies suggested that a value of $C_{f} \approx 0.3$ is sufficient for most suspension applications.

During complete opening and closing of the shim stack or valve, an impact force
\begin{equation}
F_{i,j}\left(x_{j}\right)=\begin{cases}
-k_{i}\left(x_{j}-x_{d}\right), & \quad x_{j} \leq x_{d},\\
0, & \quad x_{d}<x_{j}<x_{u},\\
-k_{i}\left(x_{j}-x_{u}\right), & \quad x_{j} \geq x_{u},
\end{cases}
\label{Eq:2B-2-3} 
\end{equation}
acts on the plates that prohibits a further motion against its resistance.
It resembles an elastic recoil force of a stiffness $k_{i}$, much larger than any system-inherent stiffness (i.e. $k_{i} \gg k_{j}$, $j=1,2$). 
Here, the values $x_{d} \approx 0$ and $x_{u} \sim \mathcal{O}(10^{-3})$ are lower and upper bounds to the shim motion given by geometric limitations.

Finally, a pretension force $F_{0}$ is sometimes applied to the valve that allows for targeted delay of its opening.
Together with an inertial term $m_{j} \ddot{x}$, which accounts for the additional acceleration $\ddot{x}$ by the damper rod, a summation of the above forces following Newton's second law gives an equation of motion for each shim stack and valve $j$ in the shock absorber system.
It is of the form
\begin{equation}
m_{j} \ddot{x}_{j} = - c_{j}\dot{x}_{j} - k_{j}x_{j} + F_{p,j} + F_{m,j} + F_{i,j} - F_{0,j} + m_{j}\ddot{x}.
\label{Eq:2B-2-4}
\end{equation}

\subsubsection{Treatment of shim stack stiffness}
\label{Sec:2B-3}

\begin{figure*}[htbp]
\begin{centering}
\begin{overpic}[width=0.95\textwidth]{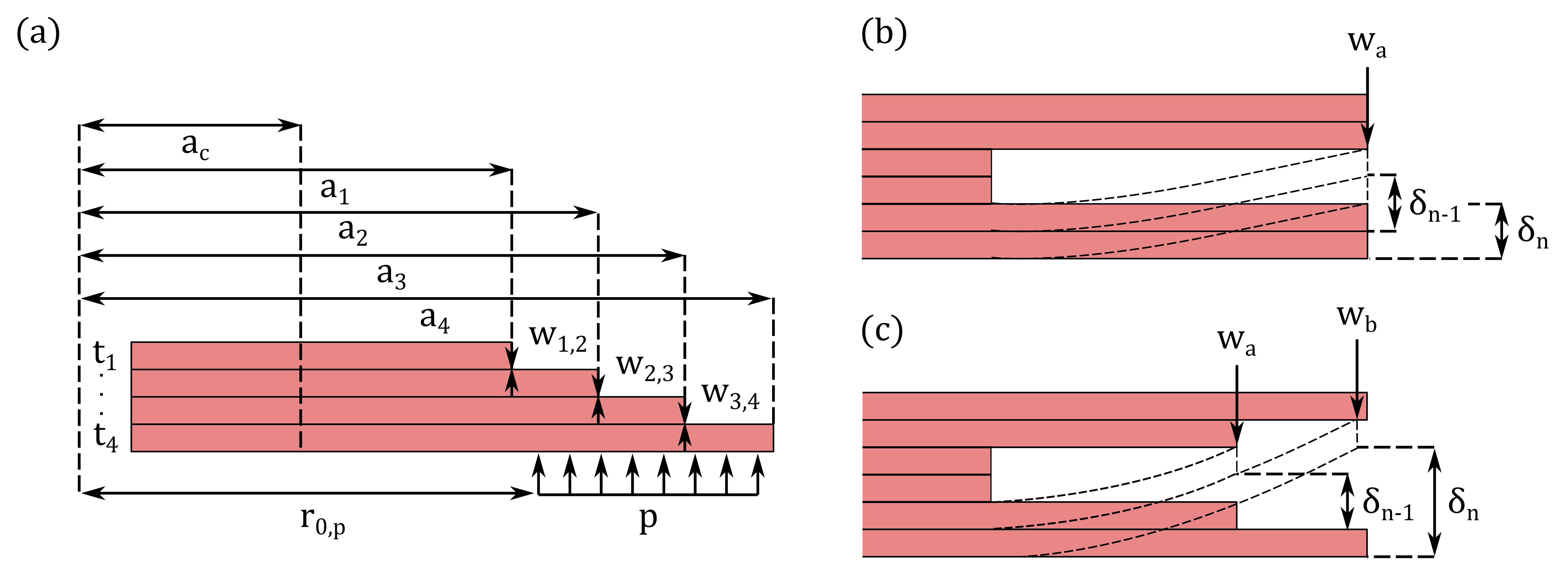}
\end{overpic}
\par
\end{centering}
\caption{Diagram of the characteristic scales of a shim stack of four shims ($n=4$) with indication of the shim radii $a_{k}$, as well as uniform pressure $p$ and line loads $w_{l,k}$ (a). 
The shim stack stiffness is computed by considering the force balance across each annular shim and by computing all resulting deformations, as outlined in Sec. \ref{Sec:2B-3}.
The specific case of crossover shims leads to additional stiffness when shims are touching, while the total deflection $\delta$ of a shim stack results from the final shim deflection $\delta_{n}$ (b). Additional nonlinearities of the shim stack stiffness can be introduced by using intermediate crossover shims (c). The numerical treatment of crossovers is outlined in Appendix \ref{Sec:A3}.}
\label{Fig:2B-3-1} 
\end{figure*}

Since the deflection of a shim stack $\delta$ occurring in the shock absorber is typically of the same order as the shim thickness $t_{k}$, $\mathcal{O}\left(\delta\right) \approx \mathcal{O}\left(t_{k}\right)$, a linearisation of the mechanical deformation can be performed.
Therefore, the total stiffness of each shim stack is calculated through the method of superposition. It uses a linear summation of deflections due to single loads on each shim in order to include all boundary conditions. By taking into account both uniform pressure loads from the fluid, as well as the initially unknown reaction forces between the shims, an expression for the deflection of each shim can be obtained. 
A regular shim stack, as shown in Fig. \ref{Fig:2B-3-1}(a), is composed of annular shims of increasing radii $a_{k}$ from top to bottom that are arranged in pyramid form. 
For the computation of the total stiffness $k_{tot}$ of a shim stack to be used for the equation of motion of driven-harmonic-oscillator type as given by Eq. (\ref{Eq:2B-2-4}),
the following assumptions are applied:
\begin{itemize}
\item The pressure load is assumed to act uniformly from a radius $r_{0,p}$ to the outer radius $a_{n}$ of the last shim of a stack of $n$ shims. In reality the pressure distribution follows a sharp parabolic profile, but the gradient at the bounds of the area of acting pressure $A_{p}$ is very steep and within those bounds mostly homogeneous, such that uniformity is a reasonable approximation.
\item The last shim then acts on the previous one via a line load across its radius. Indeed, the majority of forces between the deflected successive shims is transferred across their outer bounds.
An equal reaction force is assumed to work in the opposite direction on the lower shim, as depicted in Fig. \ref{Fig:2B-3-1}(a).
\item By summing up the displacements due to each force acting on a single shim, its deflection $\delta_{k}$ is computed according to the method of superposition. 
Furthermore, the total deflection $\delta$, which is the resulting opening $x_{j}$ of the shim stack $j$, is then equivalent to the deflection of the last shim $\delta_{n}$, as indicated in Fig. \ref{Fig:2B-3-1}(b).
\end{itemize} 
The above assumptions are applied to a shim stack of $n$ shims as follows:
\begin{align}
\delta_{1} &= \left(\delta_{1}\right)_{w_{2,1}},  \label{Eq:2B-3-1a}  \\
\delta_{2} &= \left(\delta_{2}\right)_{w_{1,2}} + \left(\delta_{2}\right)_{w_{3,2}},  \label{Eq:2B-3-1b}  \\
&\vdotswithin{=} \notag \\
\delta_{n-1} &= \left(\delta_{n-1}\right)_{w_{n-2,n-1}} + \left(\delta_{n-1}\right)_{w_{n,n-1}},  \label{Eq:2B-3-1c}  \\
\delta_{n} &= \left(\delta_{n}\right)_{w_{n-1,n}} + \left(\delta_{n}\right)_{q}.  \label{Eq:2B-3-1d}  
\end{align}
Here, the deflection $\left(\delta_{k}\right)_{w}$ is solely due to the line load $w$, while for the contact forces between the shims the equivalence $w_{l}\vcentcolon=w_{k,l}=w_{l,k}$ applies due to Newton's third law. Furthermore, the pressure load per unit area $q$ exerted by the fluid is only acting on the last shim of Eq. (\ref{Eq:2B-3-1d}) in the example above.

The stress-strain relationship depends on the specific geometry, which in this case are annular plates fixed at a clamping radius $a_{c}$ and objected to either uniform pressure or line loads.
Therefore, the maximum edge displacement of the last shim due to the acting pressure can be written as
\begin{equation}
\left(\delta_{n}\right)_{q} = M_{rc}\frac{a_{n}^{2}}{D}C_{2} + Q_{c}\frac{a_{n}^{3}}{D}C_{3} - q\frac{a_{n}^{4}}{D}L_{11},
\label{Eq:2B-3-2}
\end{equation}
with the load per unit area $q \vcentcolon= \Delta p = F_{p}/A_{p}$, the shim radius $a_{n}$, and
\begin{align}
M_{rc} &= - q\frac{a_{n}^{2}}{C_{8}} \left[\frac{C_{9}}{2a_{n}a_{c}}\left(a_{n}^{2}-r_{0,p}^{2}\right)-L_{17}\right],  \label{Eq:2B-3-3a}  \\
Q_{c} &= q\frac{1}{2a_{c}} \left(a_{n}^{2}-r_{0,p}^{2}\right),  \label{Eq:2B-3-3b}  \\
D &= \frac{Et_{n}^{3}}{12\left(1-\nu^{2}\right)},  \label{Eq:2B-3-3c}  
\end{align}
where $M_{rc}$ and $Q_{c}$ are reaction moment and load at the clamping point, while $D$ is the flexural rigidity of the plate.
The appropriate values for the integration coefficients are documented in the collection of formulas by Roark and Young \citep{roark1989formulas} and are listed in Appendix \ref{Sec:A2}.

As a result of the fluid pressure, the last shim deforms and subjects the adjacent shim to a force across its radius, which initiates a chain reaction for all subsequent shims.
The displacement of shim $k$ due to a unit line load $w_{l} \vcentcolon= F_{l}/r_{0}$ at radius $r_{0}$ is
\begin{equation}
\left(\delta_{k}\right)_{w_{l}} = - \frac{w_{l} a_{k}}{D} \left[\frac{C_{2}}{C_{8}}\left(\frac{r_{0}C_{9}}{a_{c}}-L_{9}\right)-\frac{r_{0}C_{3}}{a_{c}}+L_{3}\right],
\label{Eq:2B-3-4}
\end{equation}
with
\begin{equation}
D = \frac{Et_{k}^{3}}{12\left(1-\nu^{2}\right)}.  
\label{Eq:2B-3-5}  
\end{equation}

Using the above expressions and simplifying the notation, one arrives at a system of equations, which connects the displacements of adjacent shims:
\begin{align}
\delta_{1}\left(w_{1}\right) &= \epsilon_{2}\left(w_{1}\right)+\epsilon_{2}\left(w_{2}\right),  \label{Eq:2B-3-6a}  \\
\delta_{2}\left(w_{1}\right)+ \delta_{2}\left(w_{2}\right) &= \epsilon_{3}\left(w_{2}\right)+\epsilon_{3}\left(w_{3}\right),  \label{Eq:2B-3-6b}  \\
&\vdotswithin{=} \notag \\
\delta_{n-2}\left(w_{n-3}\right) + \delta_{n-2}\left(w_{n-2}\right) &= \epsilon_{n-1}\left(w_{n-2}\right)+\epsilon_{n-1}\left(w_{n-1}\right),  \label{Eq:2B-3-6c}  \\
\delta_{n-1}\left(w_{n-2}\right) + \delta_{n-1}\left(w_{n-1}\right) &= \epsilon_{n}\left(w_{n-1}\right)+\epsilon_{n}\left(q\right).  \label{Eq:2B-3-6d}  
\end{align}
Here, the deflection at the outer radius $\delta$ is set equal to the deflection of the next shim $\epsilon$ at the same radius.
Moreover, the deflection due to opposing forces has to be accounted for by opposite signs depending on the reference direction. 
The shim stack under these conditions is considered as a \emph{statically indeterminate system}, where the contact forces are unknown.
By rearranging the system (\ref{Eq:2B-3-6a})-(\ref{Eq:2B-3-6d}), the contact forces $\vec{w}=\left[w_{l}\right]$ between the shims are computed based on the solution of a linearized algebraic system of the form 
\begin{equation}
\hat{A}\cdot\vec{w}=\vec{b} \quad\rightarrow\quad \vec{w}=\hat{A}^{-1}\cdot\vec{b},  
\label{Eq:2B-3-7}  
\end{equation}
with
\begin{align}
  \hat{A} &=
  \left[ {\begin{array}{cccc}
    c_{\delta_{1}w_{1}}-c_{\epsilon_{2}w_{1}} & -c_{\epsilon_{2}w_{2}} & 0 & 0  \\
    c_{\delta_{2}w_{1}} & c_{\delta_{2}w_{2}}-c_{\epsilon_{3}w_{2}} & -c_{\epsilon_{3}w_{3}} & 0  \\
    0 &  \ddots & \ddots & -c_{\epsilon_{n-1}w_{n-1}} \\
   0 & 0 & c_{\delta_{n-1}w_{n-2}} & c_{\delta_{n-1}w_{n-1}}-c_{\epsilon_{n}w_{n-1}}\\
  \end{array} } \right],  \label{Eq:2B-3-8}  \\
  \vec{w} &= 
  \left[ {\begin{array}{c}
    w_{1}  \\
    w_{2}  \\
    \vdots \\
    w_{n-1} \\
  \end{array} } \right],  \label{Eq:2B-3-9}  \\
    \vec{b} &= 
  \left[ {\begin{array}{c}
    0  \\
    \vdots  \\
    0 \\
    \epsilon_{n}\left(q\right) \\
  \end{array} } \right].  \label{Eq:2B-3-10}    
\end{align}
The coefficients $c_{\delta_{k}w_{l}}$ give the linear displacement $\delta_{k}$ due to the line load $w_{l}$ as in $\delta_{k}=c_{\delta_{k}w_{l}} \cdot w_{l}$.
They follow directly from Eq. (\ref{Eq:2B-3-4}) as
\begin{equation}
c_{\delta_{k}w_{l}} = - \frac{a_{k}}{D} \left[\frac{C_{2}}{C_{8}}\left(\frac{r_{0}C_{9}}{a_{c}}-L_{9}\right)-\frac{r_{0}C_{3}}{a_{c}}+L_{3}\right].
\label{Eq:2B-3-10a}
\end{equation}
Finally, the resulting loads $\vec{w}$ can be used to compute the final displacement of each shim via
\begin{equation}
\hat{K}\cdot\vec{w}=\vec{\delta},  
\label{Eq:2B-3-11}  
\end{equation}
where the stiffness matrix is defined as
\begin{align}
  \hat{K} &=
  \left[ {\begin{array}{cccc}
    c_{\delta_{1}w_{1}} & 0 & 0 & 0  \\
    c_{\delta_{2}w_{1}} & c_{\delta_{2}w_{2}} & 0 & 0  \\
    0 &  \ddots & \ddots & 0 \\
   0 & 0 & c_{\delta_{n-1}w_{n-2}} & c_{\delta_{n-1}w_{n-1}} \\
  \end{array} } \right].  \label{Eq:2B-3-12}   
\end{align}

In order to get a reasonable estimate for the inner boundary $r_{0,p}$ of the pressure exerted on the last shim, the resulting area is set equal to the total area of acting pressure at the considered shim stack, $A_{p} = \left(a_{n}^2-r_{0,p}^2\right)\pi$, which gives
 \begin{equation}
r_{0,p} = \sqrt{a_{n}^2-\frac{A_{p}}{\pi}}.
\label{Eq:2B-3-13a} 
\end{equation}
The radius $r_{0}$ for the contact line forces crucially depends on the form of the shim stack, i.e. if the radius of the acting shim $a_{l}$ is smaller or larger than the radius of the reacting one $a_{k}$, such that
\begin{equation}
r_{0} = \begin{cases}
a_{k}, & \quad a_{l}>a_{k}, \\
a_{l}, & \quad a_{l}<a_{k}. \\
\end{cases}
\label{Eq:2B-3-13b} 
\end{equation}
The second case might occur with shim stacks involving \emph{crossover shims}, such as the one pictured in Fig. \ref{Fig:2B-3-1}(b)-(c).
Crossovers are used in suspension applications to introduce a discontinuity to the total stack stiffness, where the lower part of the stack easily deforms with small stiffness until contact with the upper part of the stack is reached and the stiffness suddenly increases.
Then the total stiffness of the stack is influenced in a discontinuous way by the contact between lower and upper shims, which happens during opening.
Such a case can also be considered in the model by checking the shim displacements and accounting for additional contact forces in the flexibility and stiffness matrices.
The modelling approach for crossover shim stacks is described in detail in Appendix \ref{Sec:A3}.
Fig. \ref{Fig:2B-3-2} shows a comparison between a regular shim stack and one with crossover.
It can be clearly recognised that the crossover shim introduces an offset in displacement between its neighbouring shims of exactly its thickness, such that $\delta_{15}-\delta_{13}=t_{14}$.

\begin{figure*}[htbp]
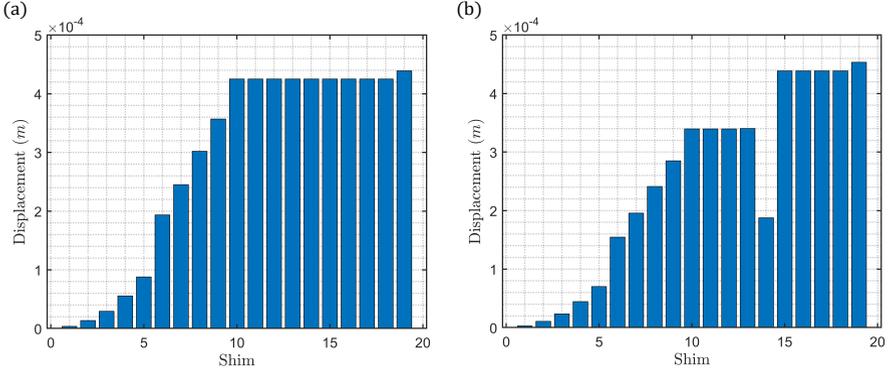

\begin{centering}
\begin{overpic}[width=1.00\textwidth]{Figure6.pdf}
\end{overpic}
\par
\end{centering}
\caption{Shim displacements for the compression stack of the baseline setup given in Appendix \ref{Sec:A1} at a fluid pressure of \SI{1}{\mega\pascal}.
The case with a regular shim stack in pyramid form and $a_{14}=a_{13}$ (a) is compared to the original case with shim no. 14 as crossover shim and $a_{14}<a_{13}$ (b).}
\label{Fig:2B-3-2} 
\end{figure*}

By using the computed partial displacements $\delta_{k}$ of all $n$ shims, which are themselves the sum of displacements due to each force acting on a single shim $k$ according to the method of superposition, the effective total stiffness of the shim stack $j$ can be determined as
\begin{equation}
k_{j} \equalhat k_{tot} = \frac{F_{tot}}{\delta},  
\label{Eq:2B-3-14}  
\end{equation}
where $F_{tot}$ is the total force acting on the shim stack, which is essentially the sum of the pressure force $F_{p,j}$ and possible opposing pretension force $F_{0,j}$, and $\delta$ is the resulting displacement of the shim stack equivalent to the displacement $\delta_{n}$ of the last shim.

\subsubsection{Calculation of damping characteristics}
\label{Sec:2B-4}

In practice, the performance of shock absorbers is frequently tuned by observing the damping force as a function of rod displacement or velocity.
The rod takes up the excitation due to road irregularities and is pushed inside the damper compartment.
The excitation is often modelled as a sinusoidal function of the form
\begin{equation}
x\left(t\right) = x_{0}\sin\left(2\pi f t - \frac{\pi}{2}\right),  
\label{Eq:2B-4-1}  
\end{equation}
which is also applied in test bench measurements.

Thus, also rod velocity $\dot{x}$ and acceleration $\ddot{x}$ are obtained directly as analytic functions of the excitation frequency $f$ and the amplitude $x_{0}$ and can be accounted for in the above equations for valve motion and pressure changes.
Here, a typical amplitude of $x_{0} = \SI{0.02}{\meter}$ and a frequency of $f \approx \SI{8}{\hertz}$ are applied as a starting point, but higher frequencies are commonly used to test for hysteretic behaviour or instabilities.

Eventually, the damping force can be computed from the variables of the dampers in Fig. \ref{Fig:2A-1} as 
\begin{equation}
F_{d} = - p_{1}A_{1} + p_{2}A_{2} + \sign \left(p_{1}-p_{2}\right) F_{f} + m_{tot}\ddot{x}, 
\label{Eq:2B-4-2}  
\end{equation}
with the seal friction of the main piston $F_{f}$, and the total mass of the moving piston and rod $m_{tot} = m_{piston} + m_{rod}$.

\subsubsection{Material properties}
\label{Sec:2B-5}

The shock absorber system and in particular the mineral oil have temperature-dependent properties, which affect damper characteristics.

The dynamic viscosity of the oil can be represented by the Guzmann-Carrancio equation
\begin{equation}
\mu\left(T\right) = \mu_{0} e^{\frac{E_{a}}{RT}} \equalhat \mu_{0} e^{\frac{E_{a,0}}{R\left(T-T_{0}\right)}},  
\label{Eq:2B-5-1}  
\end{equation}
with the characteristic activation energy of oil molecules $E_{a}$, the universal gas constant $R$, and the constant viscosity of oil $\mu_{0}$ at reference temperature $T=T_{0}$ \citep{haj2014contribution}.
This can be expressed as 
\begin{equation}
\mu\left(T\right) \approx \mu_{0} e^{C \left(\frac{1}{T}-\frac{1}{T_{0}}\right)}. 
\label{Eq:2B-5-2}  
\end{equation}
Dixon \citep{dixon2008shock} states that a reasonable estimate for the temperature sensitivity coefficient $C$ of regular unimproved mineral oils for dampers is
\begin{equation}
C = 5693 - 304\log_{10}\left(\mu_{15}\right) - 646\log_{10}^{2}\left(\mu_{15}\right),  \nonumber
\label{Eq:2B-5-3}  
\end{equation}
where $\mu_{15}$ is the measured viscosity at \SI{15}{\celsius}, applicable for a range of $0.003 < \mu_{15} < \SI{0.300}{\pascal\second}$.
By using representative values for light damper oil such as $\mu_{15} = \mu\left(T=\SI{288}{\kelvin}\right) = \SI{10}{\milli\pascal\second}$ with $\log_{10}{\mu_{15}}=-2$, this leads to a temperature sensitivity of $C = \SI{3717}{\kelvin}$.
In case of improvements by additives such as silicon, the oil's sensitivity might be decreased by up to a half. 

Furthermore, the temperature-dependent density of the damper oil can be written as
\begin{equation}
\rho\left(T\right) = \rho_{0} \frac{1}{1+\alpha_{T}\left(T-T_{0}\right)},  
\label{Eq:2B-5-4}  
\end{equation}
with the coefficient of volumetric thermal expansion $\alpha_{T}$ and the density $\rho_{0}$ at reference temperature $T=T_{0}$.
This expression gives a good accuracy for compressible oils over a wide range of temperatures with a cubic thermal expansion of $\alpha_{T} \approx \SI{0.001}{\per\kelvin}$ and a reference density of $\rho_{0} = \rho\left(T=\SI{288}{\kelvin}\right) = \SI{830}{\kilogram\per\cubic\meter}$.

Finally, the compressibility of the oil is assumed constant, but an additional offset is induced by the small elasticity of the shock absorber compartment \citep{lang1977study}.
Therefore, the resulting effective compressibility is given by 
\begin{equation}
\beta = \beta_{0} + \frac{2 r_{d}}{E_{d} t_{d}},  
\label{Eq:2B-5-5}  
\end{equation}
with the radius $r_{d}$ and the wall thickness $t_{d}$ of the damper cylinder, as well as the Young's modulus $E_{d} = \SI{200e9}{\pascal}$ of the stainless steel of the compartment.
The baseline oil compressibility is taken as $\beta_{0} = \SI{6.6e-10}{\per\pascal}$.

\subsection{Monotube system}
\label{Sec:2C}

By writing Eq. (\ref{Eq:2B-1-11a})-(\ref{Eq:2B-1-12}) for the pressure gradients within the chambers and Eq. (\ref{Eq:2B-2-4}) for the motion of each valve of the monotube shock absorber as shown in Fig. \ref{Fig:2A-1}(a), one obtains a system of first-order, nonlinear, coupled equations:
\begin{align}
\dot{y}_{1} &= y_{4},  \label{Eq:2C-1}  \\
\dot{y}_{2} &= y_{5},  \label{Eq:2C-2}  \\
\dot{y}_{3} &= y_{6},  \label{Eq:2C-3}  \\
\dot{y}_{4} &= \frac{1}{m_{1}} \left( - c_{1}y_{5} - k_{1}y_{1} + F_{p,1} + F_{m,1} + F_{i,1} - F_{0,1} + m_{1}\ddot{x} \right),  \label{Eq:2C-4}  \\
\dot{y}_{5} &= \frac{1}{m_{2}} \left( - c_{2}y_{6} - k_{2}y_{2} + F_{p,2} + F_{m,2} + F_{i,2} - F_{0,2} - m_{2}\ddot{x} \right),  \label{Eq:2C-5}  \\
\dot{y}_{6} &= \frac{1}{m_{g}} \left[ A_{g}\left(y_{7}-y_{9}\right) - F_{f,g} \right],  \label{Eq:2C-6}  \\
\dot{y}_{7} &= \frac{1}{\beta \left[A_{2}\left(L_{2}-x\right)+A_{g}y_{3}\right]} \left( - Q_{v,1} - Q_{v,2} - Q_{b} - Q_{l} + A_{2}\dot{x} - A_{g}y_{6} \right),  \label{Eq:2C-7}  \\
\dot{y}_{8} &= \frac{1}{\beta \left[A_{1}\left(L_{1}+x\right)\right]} \left( Q_{v,1} + Q_{v,2} + Q_{b} + Q_{l} - A_{1}\dot{x} \right),  \label{Eq:2C-8}  \\
\dot{y}_{9} &= - \gamma \frac{y_{9}\left(-A_{g}y_{6}\right)}{A_{g}\left(L_{g}-y_{3}\right)}.  \label{Eq:2C-9} 
\end{align}
The system is coupled in the sense that its variables, which are defined as
\begin{align}
    \vec{y} &= \begin{bmatrix}
              y_{1} \\
              y_{2} \\
              y_{3} \\
              y_{4} \\
              y_{5} \\
              y_{6} \\
              y_{7} \\
              y_{8} \\             
              y_{9}                 
             \end{bmatrix} =
              \begin{bmatrix}
              x_{1} \\
              x_{2} \\
              x_{g} \\
              \dot{x}_{1} \\
              \dot{x}_{2} \\
              \dot{x}_{g} \\
              p_{2} \\
              p_{1} \\             
              p_{g}                 
             \end{bmatrix} \equalhat
             \begin{bmatrix}
              x_{c} \\
              x_{r} \\
              x_{g} \\
              \dot{x}_{c} \\
              \dot{x}_{r} \\
              \dot{x}_{g} \\
              p_{c} \\
              p_{r} \\             
              p_{g}                 
             \end{bmatrix},
\label{Eq:2C-10}             
\end{align}
crucially depend on each other, since the valve and gas piston displacements $x_{1},x_{2}, x_{g}$ are impacted by the pressures $p_{1},p_{2},p_{g}$, and vice versa. Here, the quantities refer to the notation as used in Fig. \ref{Fig:2A-1}(a), which further equate to the deflection of the shim valve during compression $x_{c}$, and of the one during rebound $x_{r}$.

The system of Eq. (\ref{Eq:2C-1})-(\ref{Eq:2C-9}) can be written in compact form as
\begin{equation}
\dot{\vec{y}}=\vec{f}\left(\vec{y}\right),  \nonumber
\label{Eq:2C-11}
\end{equation}
and subsequently solved with common numerical schemes as described in Sec. \ref{Sec:2E}.

\subsection{Piggyback system}
\label{Sec:2D}

It can be seen in Fig. \ref{Fig:2A-1} that the only crucial difference between the two considered types of shock absorbers is the additional fixed check valve piston of the piggyback damper. As a result, their analytical models are similar, except for the equations of motion for the two additional valves and the balance equation for the pressure in the second compression chamber \circled{3}. Consequently, the full equation system of the piggyback geometry reads as

\begin{align}
\dot{y}_{1} &= y_{6},  \label{Eq:2D-1}  \\
\dot{y}_{2} &= y_{7},  \label{Eq:2D-2}  \\
\dot{y}_{3} &= y_{8},  \label{Eq:2D-3}  \\
\dot{y}_{4} &= y_{9},  \label{Eq:2D-4}  \\
\dot{y}_{5} &= y_{10},  \label{Eq:2D-5}  \\
\dot{y}_{6} &= \frac{1}{m_{1}} \left( - c_{1}y_{6} - k_{1}y_{1} + F_{p,1} + F_{m,1} + F_{i,1} - F_{0,1} + m_{1}\ddot{x} \right),  \label{Eq:2D-6}  \\
\dot{y}_{7} &= \frac{1}{m_{2}} \left( - c_{2}y_{7} - k_{2}y_{2} + F_{p,2} + F_{m,2} + F_{i,2} - F_{0,2} - m_{2}\ddot{x} \right),  \label{Eq:2D-7}  \\
\dot{y}_{8} &= \frac{1}{m_{3}} \left( - c_{1}y_{8} - k_{3}y_{3} + F_{p,3} + F_{m,3} + F_{i,3} - F_{0,3} \right),  \label{Eq:2D-8}  \\
\dot{y}_{9} &= \frac{1}{m_{4}} \left( - c_{2}y_{9} - k_{4}y_{4} + F_{p,4} + F_{m,4} + F_{i,4} - F_{0,4} \right),  \label{Eq:2D-9}  \\
\dot{y}_{10} &= \frac{1}{m_{g}} \left[ A_{g}\left(y_{13}-y_{14}\right) - F_{f,g} \right],  \label{Eq:2D-10}  \\
\dot{y}_{11} &= \frac{1}{\beta \left[A_{2}\left(L_{2}-x\right)\right]} \left( - Q_{v,1} - Q_{v,2} - Q_{b} - Q_{l} + A_{2}\dot{x} \right),  \label{Eq:2D-11}  \\
\dot{y}_{12} &= \frac{1}{\beta \left[A_{1}\left(L_{1}+x\right)\right]} \left( Q_{v,1} + Q_{v,2} + Q_{b} + Q_{l} - A_{1}\dot{x} \right),  \label{Eq:2D-12}  \\
\dot{y}_{13} &= \frac{1}{\beta \left[A_{3}\left(L_{3}-x\right)+A_{g}y_{5}\right]} \left( Q_{v,3} + Q_{v,4} + \tilde{Q}_{b} - A_{g}y_{10} \right),  \label{Eq:2D-13}  \\
\dot{y}_{14} &= - \gamma \frac{y_{14}\left(-A_{g}y_{10}\right)}{A_{g}\left(L_{g}-y_{5}\right)},  \label{Eq:2D-14} 
\end{align}
with the solution vector
\begin{align}
    \vec{y} &= \begin{bmatrix}
              y_{1} \\
              y_{2} \\
              y_{3} \\
              y_{4} \\
              y_{5} \\
              y_{6} \\
              y_{7} \\
              y_{8} \\             
              y_{9} \\          
              y_{10} \\
              y_{11} \\
              y_{12} \\
              y_{13} \\             
              y_{14}          
             \end{bmatrix} =
             \begin{bmatrix}
              x_{1} \\
              x_{2} \\
              x_{3} \\
              x_{4} \\              
              x_{g} \\
              \dot{x}_{1} \\
              \dot{x}_{2} \\
              \dot{x}_{3} \\
              \dot{x}_{4} \\
              \dot{x}_{g} \\
              p_{2} \\
              p_{1} \\
              p_{3} \\             
              p_{g}                 
             \end{bmatrix} \equalhat
             \begin{bmatrix}
              x_{c} \\
              x_{r} \\
              \tilde{x}_{c} \\
              \tilde{x}_{r} \\              
              x_{g} \\
              \dot{x}_{c} \\
              \dot{x}_{r} \\
              \dot{\tilde{x}}_{c} \\
              \dot{\tilde{x}}_{r} \\
              \dot{x}_{g} \\
              p_{c} \\
              p_{r} \\
              \tilde{p}_{c} \\             
              p_{g}                 
             \end{bmatrix},
\label{Eq:2D-15}             
\end{align}
where the notation follows the sketch of Fig. \ref{Fig:2A-1}(b), and the displacements of the valves on the additional piston $\tilde{x}_{c},\tilde{x}_{r}$ and the additional compression chamber pressure $\tilde{p}_{c}$ are marked with a tilde.

\subsection{Numerical method}
\label{Sec:2E}

Both systems outlined in Sec. \ref{Sec:2C}-\ref{Sec:2D} are comprised of \emph{stiff equations}, where some variables might show vanishingly small gradients, while others change drastically. Here, an example would be during the compression stage, where the rebound valves would be closed, while the compression valves open during rapid variation of the rebound and compression pressures that are several orders of magnitude larger.
Although general explicit integration schemes such as 4\textsuperscript{th}-order Runge-Kutta with a local truncation error of $\mathcal{O}(\Delta t^{4})$ might work under certain conditions, the time step would have to be forced to very small values to meet the assigned tolerances, which would lead to long overall computation times.
Therefore,  implicit schemes of the boundary differentiation formulation (BDF) are applied, which have superior performance and numerical stability with regards to stiff equation systems, such as the one in this study.
Overall, BDF has been shown to need considerably less computation time at similar tolerances than Runge-Kutta methods \citep{stewart1990avoiding}.
Another powerful option are hybrid integration schemes such as LSODE that essentially combine regular explicit Runge-Kutta with implicit BDF schemes \citep{hindmarsh1983odepack,radhakrishnan1993description}.

Suitable LSODE/LSODA/BDF libraries for the numerical solution of the systems presented here can be applied through the free scientific programming languages \emph{GNU Octave} or \emph{Python}. Additionally, implicit solvers for such stiff equations within \emph{MATLAB} are for instance ODE15s or ODE23s.
As an example, ODE15s at a maximum order of 5, with adaptive timestepping and relative and absolute error tolerances of $\mathcal{O}(10^{-4})$ leads to computation times of consistently below \SI{30}{\second} for both systems of Sec. \ref{Sec:2C}-\ref{Sec:2D}, often even as low as \SI{5}{\second}.


\section{Validation}
\label{Sec:3}

\begin{figure*}[htbp]
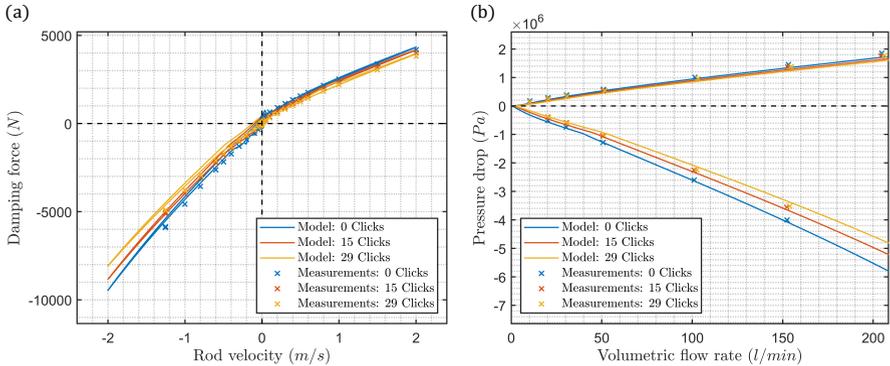

\begin{centering}
\begin{overpic}[width=1.00\textwidth]{Figure7.pdf}
\end{overpic}
\par
\end{centering}
\caption{Comparison of damping curves (a) and pressure drop across the main piston (b) obtained from the analytical model and from test bench measurements of a piggyback damper for bleed flow openings of \SI{0}{\clicks} ($A_{b}=\SI{0}{\square\milli\meter}$, $\tilde{A}_{b}=\SI{0}{\square\milli\meter}$), \SI{15}{Clicks} ($A_{b}=\SI{3.9}{\square\milli\meter}$, $\tilde{A}_{b}=\SI{1.4}{\square\milli\meter}$) and \SI{29}{\clicks} ($A_{b}=\SI{7.54}{\square\milli\meter}$, $\tilde{A}_{b}=\SI{2.7}{\square\milli\meter}$).}
\label{Fig:3-1} 
\end{figure*}

The model presented in Sec. \ref{Sec:2} is validated for the case of a piggyback damper based on the system properties given in Appendix \ref{Sec:A1}.
Test bench data is obtained for the cases of $0$, $15$, and \SI{29}{\clicks}, where \SI{}{\clicks} refers to a common configuration unit for the adjustment of the bleed flow cross-sectional area of the shock absorber. Here, this relates to bleed orifice areas of main and check valve piston of $A_{b}=\SI{0}{\square\milli\meter}$, $\tilde{A}_{b}=\SI{0}{\square\milli\meter}$  at \SI{0}{\clicks}, $A_{b}=\SI{3.9}{\square\milli\meter}$, $\tilde{A}_{b}=\SI{1.4}{\square\milli\meter}$  at \SI{15}{\clicks}, and $A_{b}=\SI{7.54}{\square\milli\meter}$, $\tilde{A}_{b}=\SI{2.7}{\square\milli\meter}$  at \SI{29}{\clicks}.
An overall very good agreement is reached and the relative deviation of the analytical prediction to the measurement data is consistently below \SI{5}{\percent} for the damping characteristics in Fig. \ref{Fig:3-1}(a) and below \SI{4}{\percent} for the main piston pressure drop in Fig. \ref{Fig:3-1}(b).

\section{Results}
\label{Sec:4}

Using the baseline properties of Appendix \ref{Sec:A1}, the models of the monotube system of Sec. \ref{Sec:2C} and of the piggyback system of Sec. \ref{Sec:2D} are applied to calculate the system variables presented below.
There are important differences in the function and inherent co-dependencies of model parameters between the two considered types of shock absorbers, which are further discussed in the following.

\subsection{Monotube shock absorber variables}
\label{Sec:4A}

\begin{figure*}[htbp]
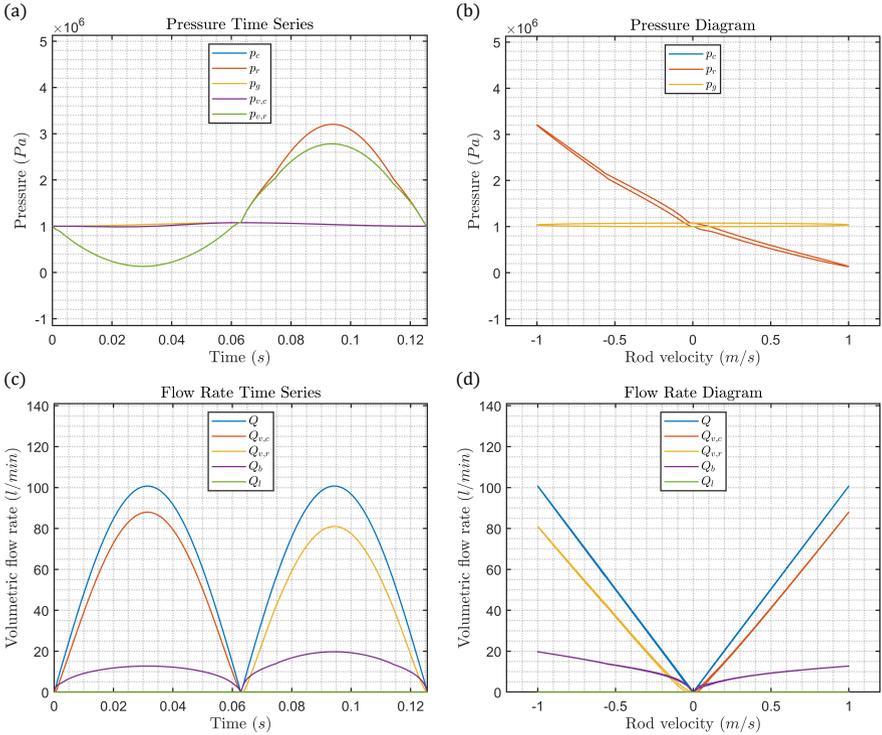

\begin{centering}
\begin{overpic}[width=1.00\textwidth]{Figure8.pdf}
\end{overpic}
\par
\end{centering}
\caption{Pressure (a)-(b) and volumetric flow rate (c)-(d) with respect to time $t$ and rod velocity $\dot{x}$ for the monotube system presented in Sec. \ref{Sec:2C}. The system variables are shown for one entire cycle consisting of compression and rebound stroke as introduced by the boundary condition of Eq. (\ref{Eq:2B-4-1}).}
\label{Fig:4A-1} 
\end{figure*}

The monotube damper consists of a piston, which separates two chamber volumes filled with oil. The pressures within the compression and rebound chamber $p_{c}$ and $p_{r}$, as well as those within the piston valve ports $p_{v,j}$ are plotted with respect to time and piston rod velocity $\dot{x}$ in Fig. \ref{Fig:4A-1}(a)-(b).
Due to the effect of the gas reservoir and its almost frictionless gas piston, which is adjacent to the compression chamber, any volumetric changes are well absorbed. This leads to an almost equal pressure in the compression chamber and adjacent gas volume throughout the entire cycle, while the pressure gradient across the piston is dominated by pressure changes within the rebound chamber (cf. Fig. \ref{Fig:4A-1}(b)).
This large variation of the rebound chamber pressure within the monotube damper makes is also fairly vulnerable to cavitation, if the pressure during the rebound stroke drops too low and the reference pressure adjusted by the gas reservoir is not chosen accordingly.
Furthermore, depending on shim stack stiffness on each side of the piston, the flow rate is distributed differently accross bleed and valve orifices during the initial compression and subsequent rebound of the shock absorber.

\begin{figure*}[htbp]
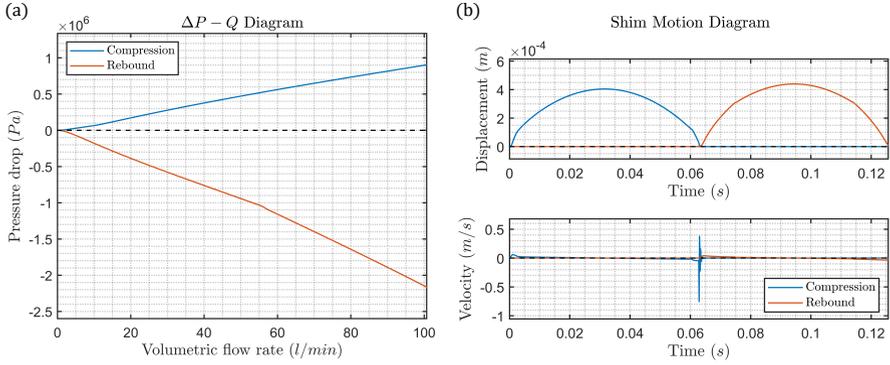

\begin{centering}
\begin{overpic}[width=1.00\textwidth]{Figure9.pdf}
\end{overpic}
\par
\end{centering}
\caption{Pressure drop across the piston (a) and motion diagram of the compression and rebound shim stack (b), showing shim displacement $x_{j}$ and velocity $\dot{x}_{j}$ of stack $j$.}
\label{Fig:4A-2} 
\end{figure*}

\begin{figure*}[htbp]
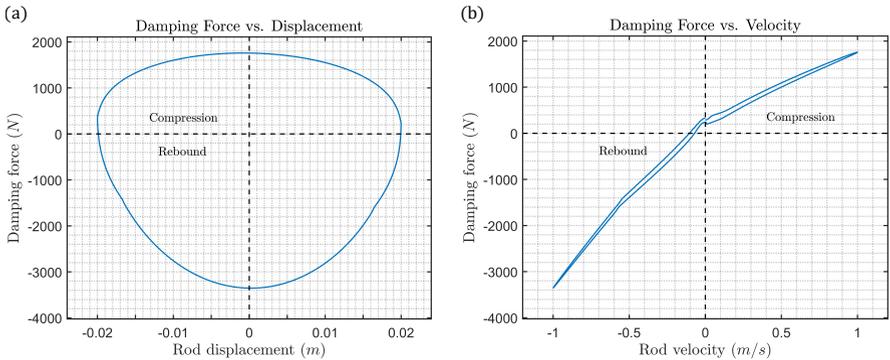

\begin{centering}
\begin{overpic}[width=1.00\textwidth]{Figure10.pdf}
\end{overpic}
\par
\end{centering}
\caption{Damping characteristics of the monotube system throughout the entire cycle as a function of rod displacement $x$ and velocity $\dot{x}$.}
\label{Fig:4A-3} 
\end{figure*}

Fig. \ref{Fig:4A-2}(a) shows the pressure drop across the piston during the compression and rebound phase. Discontinuities in the curves are introduced by the crossover shims of the compression and rebound stacks. However, this effect is particularly visible for the rebound stroke at a volume flow of approximately \SI{55}{\liter\per\min} due to the double crossover of the rebound stack, which leads to a sudden increase in stiffness at that point (cf. Appendix \ref{Sec:A1}).
The discontinuous shim stack stiffness can also be seen in the rebound damping curves of Fig. \ref{Fig:4A-3}.
The opening and closing motion of the shim stacks can be tracked by the time series in Fig. \ref{Fig:4A-2}(b). The velocities $\dot{x}_{j}$ thereby give an indication of sudden impact forces that are unwanted for improved durability of the valve.

\subsection{Piggyback shock absorber variables}
\label{Sec:4B}

\begin{figure*}[htbp]
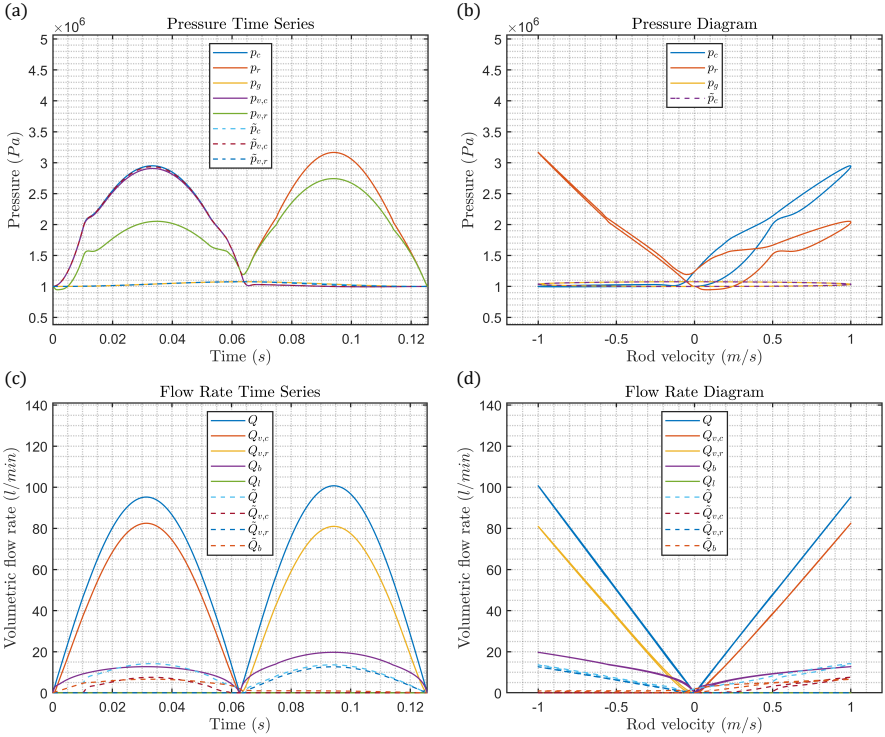

\begin{centering}
\begin{overpic}[width=1.00\textwidth]{Figure11.pdf}
\end{overpic}
\par
\end{centering}
\caption{Time series and diagram of all occurring pressures (a)-(b), as well as flow rates (c)-(d) of the piggyback system of Sec. \ref{Sec:2D}. The dashed lines and tilde mark the additional check valve piston properties and associated chamber pressures as introduced in Eq. (\ref{Eq:2D-15}).}
\label{Fig:4B-1} 
\end{figure*}

\begin{figure*}[htbp]
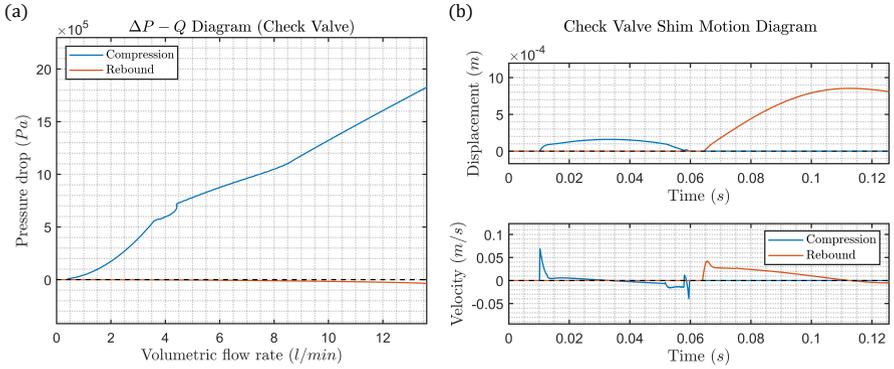

\begin{centering}
\begin{overpic}[width=1.00\textwidth]{Figure13.pdf}
\end{overpic}
\par
\end{centering}
\caption{Pressure gradient (a), as well as shim stack displacement and velocity (b) of the check valve piston of the piggyback damper.}
\label{Fig:4B-3} 
\end{figure*}

By addition of a check valve piston controlling the flow direction between main compartment and gas reservoir, the piggyback system (cf. Sec. \ref{Sec:2D}) is obtained.
The piggyback damper crucially produces pressure losses by shim deformation during the engagement of the second piston, as demonstrated in Fig. \ref{Fig:4B-3}.
This adds complexity to the system by nonlinear distribution of pressures between the three oil-filled chambers. In particular, it prohibits the rebound pressure from decreasing considerably below the reference pressure of the system, which can be seen by comparison of Fig. \ref{Fig:4B-1}(a)-(b) with Fig. \ref{Fig:4A-1}(a)-(b) for the monotube damper.
Keeping the pressure from dropping too low has the additional advantage of avoiding excessive cavitation in the shock absorber, particularly at the high-speed valve exits.
Furthermore, the dynamic behaviour of the valves can be controlled by added pretension and spring stiffness, as it is the case for the compression shim stack on the check valve piston of the analysed case (cf. Appendix \ref{Sec:A1}). The delay in shim opening visible in Fig. \ref{Fig:4B-3}(b) is due to the pretension and has direct implication on the associated flow rates $\tilde{Q}_{v,c}$ and $\tilde{Q}$ in Fig. \ref{Fig:4B-1}(c)-(d).
It is a generally unwanted effect of the studied setup that the inertia of the check valve causes it to stay open during the end phase of the rebound (cf. Fig. \ref{Fig:4B-3}). This can be alleviated by adjusting the values of the spring preload and stiffness, which are given in Appendix \ref{Sec:A1}. Thus, also the response time, which indicates the time lag of the valve towards rapid changes in flow direction, such as with a sudden compression stroke, can be crucially reduced.

\begin{figure*}[htbp]
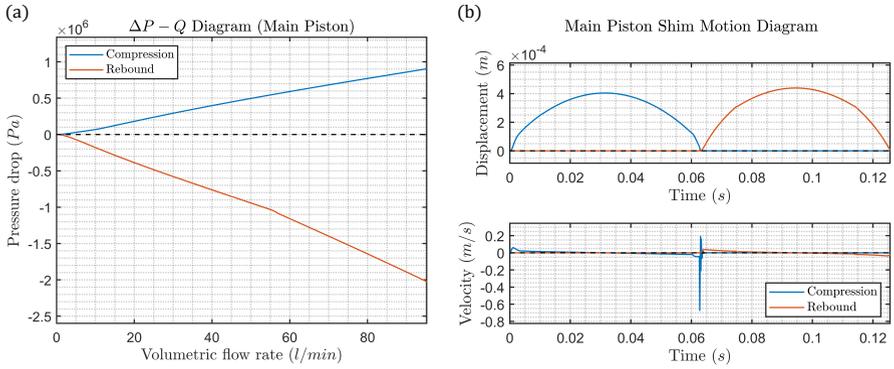

\begin{centering}
\begin{overpic}[width=1.00\textwidth]{Figure12.pdf}
\end{overpic}
\par
\end{centering}
\caption{Pressure gradient (a) and shim stack motion variables of displacement and velocity (b) of the main piston of the piggyback damper.}
\label{Fig:4B-2} 
\end{figure*}

\begin{figure*}[htbp]
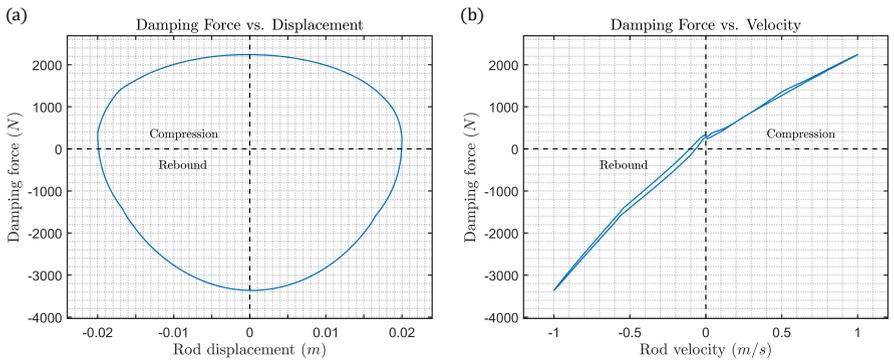

\begin{centering}
\begin{overpic}[width=1.00\textwidth]{Figure14.pdf}
\end{overpic}
\par
\end{centering}
\caption{Damping curves with respect to displacement (a) and velocity (b) of the piggyback shock absorber rod.}
\label{Fig:4B-4} 
\end{figure*}

It is evident from comparison of Fig. \ref{Fig:4A-2} and Fig. \ref{Fig:4A-3} with Fig. \ref{Fig:4B-2} and Fig. \ref{Fig:4B-4} that although the pressure drop across the main piston is the same due to identical shim stacks, the resulting damping properties of the two analysed shock absorber types are different.
The piggyback damper has crucial advantages towards the more common monotube damper - such as smaller axial dimensions, less sensitivity to temperature changes due to a quasi-isolated compartment, reduced risk of cavitation, and an additional valve for compression. 
In particular the latter feature allows for more control of the suspension system and a targeted increase of the damping force to above $\SI{2200}{\newton}$ during the compression stroke, which is demonstrated in Fig. \ref{Fig:4B-4}.


\section{Conclusions}
\label{Sec:5}

Due to their intricate design and nonlinear coupling between various model variables -- such as pressure drop, constant and variable volumetric flow rates, as well as shim stack stiffness -- shock absorbers have long evaded analytical treatment. Until recently their development was based on mostly empirical methods by comparison of a multitude of setups on test tracks in a trial-and-error fashion \citep{dixon2008shock}. This approach led to a vast body of measurement data for each damper model and each manufacturer, which allowed for optimization after years of targeted research.
However, due to the nature of the underlying systems presented in Sec. \ref{Sec:2C}-\ref{Sec:2D}, which consist of nonlinear, coupled, stiff differential equations, a development of shock absorbers neglecting analytical methods is ill-advised.
This is because of the drastic effect already small parameter changes can have on the damping performance, while other variations might prove without consequences.
Due to the early work of Lang \cite{lang1977study} and later Reybrouck \cite{reybrouck1994non}, the strength of reduced-order models was slowly recognized in the suspension community.
However, although recent work showed the strong potential of mathematical modelling based on physical principles, a common validated approach was lacking that would allow for universal application of the methods for various shock absorber types \citep{farjoud2012nonlinear,skavckauskas2017development}. 
This study provides such a method and derives all relevant system variables for the common monotube shock absorber and its derivative, the piggyback shock absorber, by using typical baseline properties. Moreover, in contrast to previous semi-empirical, algebraic models that often lacked parameters with physical relevance, the model proposed here consists of a first-order system of nonlinear differential equations, which captures the rich physics of shock absorbers.
Validation of the model is achieved by comparison with test bench measurements and the analytical prediction lies consistently within \SI{5}{\percent} (cf. Sec. \ref{Sec:3}).  
Crucial effects, such as the rebound pressure dominance of the monotube damper, or the large variation of damping force by the additional check valve compression stack of the piggyback damper, are documented in Sec. \ref{Sec:4}.
Due to the modular nature of the presented approach, as demonstrated in Sec. \ref{Sec:2}, the proposed model is also of general relevance to shock absorbers of other types (e.g. dual-tube shock absorbers), since the equation system can be extended to include multiple valves and chambers in a straight-forward manner, giving it indeed universal applicability.
Finally, the underlying equations of the mathematical model are of first-order type and well-suited for numerical integration and further studies of the non-smooth dynamical systems through bifurcation diagrams, stability maps, or possible chaotic indicators.



\begin{appendices}

\section{Baseline system properties}
\label{Sec:A1}

\setcounter{equation}{0}

The properties listed in Tab. \ref{Tab:A1-1}-\ref{Tab:A1-3} for the shock absorber geometry and the damper fluid, are used as baseline for the results presented in Sec. \ref{Sec:3}.
Where other values are used, this is explicitly stated in the text.
For the shim stacks the following setup of circular annular shims is used:
\begin{itemize}
\item The compression stack is built up of shims with radii $\left[a_{k}\right]=\left[12,13,14,15,16,18,19,20,21,22,22,22,22,16,22,22,22,22,22.5\right]\times10^{-3}\,\SI{}{\meter}$ and thicknesses $\left[t_{k}\right]=[0.20,0.25,0.25,0.20,0.20,0.20,0.20,0.20,0.25,0.20,$ \linebreak $0.20,0.20,0.20,0.10,0.20,0.20,0.20,0.20,0.20]\times10^{-3}\,\SI{}{\meter}$ with a clamp radius of $a_{c}=11\times10^{-3}\,\SI{}{\meter}$.
\item The rebound stack consists of shims with radii $\left[a_{k}\right]=\left[13,14,15,16,17,18,12,14,20,20,20,20,20\right]\times10^{-3}\,\SI{}{\meter}$ and thicknesses $\left[t_{k}\right]=[0.30,0.30,0.30,0.30,0.25,0.25,0.10,0.10,0.15,0.20,0.20,0.20,0.20]\times10^{-3}\,\SI{}{\meter}$ with a clamp radius of $a_{c}=11.5\times10^{-3}\,\SI{}{\meter}$.
\item The compression stack on the check valve piston is built up of shims with radii $\left[a_{k}\right]=\left[5,7,8,9,11,11,6,11,11,11\right]\times10^{-3}\,\SI{}{\meter}$ and thicknesses $\left[t_{k}\right]=[0.20,0.20,0.20,0.20,0.20,0.20,0.10,0.20,0.20,0.20]\times10^{-3}\,\SI{}{\meter}$ with a clamp radius of $a_{c}=4\times10^{-3}\,\SI{}{\meter}$.
Moreover, an additional spring with a pretension of $F_{0}=\SI{43.2}{\newton}$ and a stiffness of $k=\SI{16000}{\newton\per\meter}$ is applied.
\item The check valve is one annular shim of an inner radius of $4\times10^{-3}\,\SI{}{\meter}$, an outer radius of $6.3\times10^{-3}\,\SI{}{\meter}$, and a thickness of $0.30\times10^{-3}\,\SI{}{\meter}$. The valve is spring-loaded with a pretension of $F_{0}=\SI{0.35}{\newton}$ and a stiffness of $k=\SI{200}{\newton\per\meter}$.
\end{itemize}

\begin{table}[htbp]
\caption{Properties of the shock absorber pistons and compartment.}
\centering
\begin{tabular}{@{}lllll@{}}
\toprule
Parameter & Symbol & Value & Unit \\
\midrule
Reservoir pressure & $p_{g}$ & $10^{6}$ & $\SI{}{\pascal}$ \\
Reservoir length & $L_{g}$ & $70.11\times10^{-3}$ & $\SI{}{\meter}$ \\
Rebound chamber length & $L_{1}$ & $23.90\times10^{-3}$ & $\SI{}{\meter}$ \\
Compression chamber length & $L_{2}$ & $157.18\times10^{-3}$ & $\SI{}{\meter}$ \\
Second compression chamber length & $L_{3}$ & $23.04\times10^{-3}$ & $\SI{}{\meter}$ \\
Gas piston diameter & $d_{g}$ & $59.78\times10^{-3}$ & $\SI{}{\meter}$ \\
Main piston diameter & $d_{piston}$ & $49.60\times10^{-3}$ & $\SI{}{\meter}$ \\
Rod diameter & $d_{rod}$ & $17.97\times10^{-3}$ & $\SI{}{\meter}$ \\
Gas piston mass & $m_{g}$ & $0.1955$ & $\SI{}{\kilogram}$ \\
Main piston and rod mass & $m_{tot}$ & $1.0980$ & $\SI{}{\kilogram}$ \\
Cylinder radius & $r_{d}$ & $25\times10^{-3}$ & $\SI{}{\meter}$ \\
Wall thickness & $t_{d}$ & $2.5\times10^{-3}$ & $\SI{}{\meter}$ \\
Stainless steel Young's modulus & $E_{d}$ & $200\times10^{9}$ & $\SI{}{\pascal}$ \\
Seal friction & $F_{f}$ & $22.25$ & $\SI{}{\newton}$ \\
Rebound discharge coefficients & $[C_{d,c},C_{d,v}]$ & $[0.34,0.5]$ & $\SI{}{-}$ \\
Compression discharge coefficients & $[C_{d,c},C_{d,v}]$ & $[0.5,0.7]$ & $\SI{}{-}$ \\
Rebound check valve discharge coefficients & $[C_{d,c},C_{d,v}]$ & $[0.6,0.6]$ & $\SI{}{-}$ \\
Compression check valve discharge coefficients & $[C_{d,c},C_{d,v}]$ & $[0.4,0.21]$ & $\SI{}{-}$ \\
Bleed orifice discharge coefficient & $C_{d,b}$ & 0.6 & $\SI{}{-}$ \\
\bottomrule
\end{tabular}
\label{Tab:A1-1}
\end{table}

\begin{table}[htbp]
\caption{Properties of the shim stacks and valves.}
\centering
\begin{tabular}{@{}lllll@{}}
\toprule
Parameter & Symbol & Value & Unit \\
\midrule
Shim steel Poisson's ratio & $\nu$ & $0.305$ & $\SI{}{-}$ \\
Shim steel Young's modulus & $E$ & $210\times10^{9}$ & $\SI{}{\pascal}$ \\
\bottomrule
\end{tabular}
\label{Tab:A1-2}
\end{table}

\begin{table}[htbp]
\caption{Properties of the damper mineral oil.}
\centering
\begin{tabular}{@{}lllll@{}}
\toprule
Parameter & Symbol & Value & Unit \\
\midrule
Reference density & $\rho_{0}$ & $830$ & $\SI{}{\kilogram\per\cubic\meter}$ \\
Reference dynamic viscosity & $\mu_{0}$ & $0.01$ & $\SI{}{\kilogram\per\meter\per\s}$ \\
Reference compressibility & $\beta_{0}$ & $6.6\times10^{-10}$ & $\SI{}{\kilogram\per\cubic\meter}$ \\
Reference temperature & $T_{0}$ & $288$ & $\SI{}{\kelvin}$ \\
Temperature sensitivity coefficient & $C$ & $3717$ & $\SI{}{\kelvin}$ \\
Thermal expansion coefficient & $\alpha_{T}$ & $0.001$ & $\SI{}{\per\kelvin}$ \\
\bottomrule
\end{tabular}
\label{Tab:A1-3}
\end{table}

\section{Integration coefficients for shim stack stiffness}
\label{Sec:A2}

\setcounter{equation}{0}

The following coefficients after Roark and Young \citep{roark1989formulas} are used for the computation of linearized deflections at a radius $r$ of a flat shim plate of outer radius $a$ as a result of pressure or contact force loads at radius $r_{0}$:
\begin{align}
C_{8} &= \frac{1}{2} \left[1+\nu+\left(1-\nu\right)\left(\frac{b}{a}\right)^{2}\right],  \label{Eq:A2-1}  \\
C_{9} &= \frac{b}{a} \left\{ \frac{1+\nu}{2}\ln{\frac{a}{b}}+\frac{1-\nu}{4}\left[1-\left(\frac{b}{a}\right)^{2}\right] \right\},  \label{Eq:A2-2}  \\
F_{2} &= \frac{1}{4} \left[1-\left(\frac{b}{r}\right)^{2}\left(1+2\ln\frac{r}{b}\right)\right],  \label{Eq:A2-3}  \\
F_{3} &= \frac{b}{4r} \left\{\left[\left(\frac{b}{r}\right)^{2}+1\right]\ln\frac{r}{b}+\left(\frac{b}{r}\right)^{2}-1\right\},  \label{Eq:A2-4}  \\
G_{3} &= \frac{r_{0}}{4r} \left\{\left[\left(\frac{r_{0}}{r}\right)^{2}+1\right]\ln\frac{r}{r_{0}}+\left(\frac{r_{0}}{r}\right)^2-1\right\}\left<r-r_{0}\right>^{0},  \label{Eq:A2-5}  \\
G_{11} &= \frac{1}{64} \left\{1+4\left(\frac{r_{0}}{r}\right)^2-5\left(\frac{r_{0}}{r}\right)^4-4\left(\frac{r_{0}}{r}\right)^2\left[2+\left(\frac{r_{0}}{r}\right)^2\right]\ln\frac{r}{r_{0}}\right\}\left<r-r_{0}\right>^{0},  \label{Eq:A2-6}  \\
L_{9} &= \frac{r_{0}}{a} \left\{ \frac{1+\nu}{2}\ln{\frac{a}{r_{0}}}+\frac{1-\nu}{4}\left[1-\left(\frac{r_{0}}{a}\right)^{2}\right] \right\},  \label{Eq:A2-7}  \\
L_{17} &= \frac{1}{4} \left\{1-\frac{1-\nu}{4}\left[1-\left(\frac{r_{0}}{a}\right)^4\right]-\left(\frac{r_{0}}{a}\right)^2\left[1+\left(1+\nu\right)\ln{\frac{a}{r_{0}}}\right]\right\}.\label{Eq:A2-8}
\end{align} 
Here, the expression $\left<r-r_{0}\right>^{0}$ is equivalent to $\left(r-r_{0}\right) \,\forall \, r>r_{0}$ and $0 \,\forall \, r \leq r_{0}$.

\section{Modelling of crossover shim contact}
\label{Sec:A3}

\setcounter{equation}{0}

For the case of an introduced crossover shim at position $m$ within a stack, the shim deflections and stiffness are initially computed in the same way as presented in Sec. \ref{Sec:2B-3} until contact is reached between the adjacent layers of the crossover. At this point the difference between the displacements of neighbouring shims equals the total thickness of the crossover, $\delta_{m+1}-\delta_{m-1}=t_{m}$, and the new contact force $w_{a}$ needs to be taken into account:
\begin{align}
\delta_{1}\left(w_{1}\right) &= \epsilon_{2}\left(w_{1}\right)+\epsilon_{2}\left(w_{2}\right),  \label{Eq:A3-1}  \\
\delta_{2}\left(w_{1}\right)+ \delta_{2}\left(w_{2}\right) &= \epsilon_{3}\left(w_{2}\right)+\epsilon_{3}\left(w_{3}\right),  \label{Eq:A3-2}  \\
&\vdotswithin{=} \notag \\
\delta_{m-2}\left(w_{m-3}\right) + \delta_{m-2}\left(w_{m-2}\right) &= \epsilon_{m-1}\left(w_{m-2}\right)+\epsilon_{m-1}\left(w_{m-1}\right) \nonumber \\
&\quad+\epsilon_{m-1}\left(w_{a}\right),  \label{Eq:A3-3}  \\
\epsilon_{m-1}^{m}\left(w_{m-2}\right)+\epsilon_{m-1}^{m}\left(w_{m-1}\right)+\epsilon_{m-1}^{m}\left(w_{a}\right) &= \delta_{m}\left(w_{m-1}\right) + \delta_{m}\left(w_{m}\right),  \label{Eq:A3-4}  \\  
\delta_{m}\left(w_{m-1}\right) + \delta_{m}\left(w_{m}\right) &= \epsilon_{m+1}\left(w_{m}\right)+\epsilon_{m+1}\left(w_{m+1}\right) \nonumber \\
&\quad+\epsilon_{m+1}\left(w_{a}\right),  \label{Eq:A3-5}  \\
\delta_{m+1}\left(w_{m}\right) + \delta_{m+1}\left(w_{m+1}\right) + \delta_{m+1}\left(w_{a}\right) &= \epsilon_{m+2}\left(w_{m+1}\right)+\epsilon_{m+2}\left(w_{m+2}\right),\label{Eq:A3-6}  \\
\delta_{m+2}\left(w_{m+1}\right) + \delta_{m+2}\left(w_{m+2}\right) &= \epsilon_{m+3}\left(w_{m+2}\right)+\epsilon_{m+3}\left(w_{m+3}\right),  \label{Eq:A3-7}  \\
&\vdotswithin{=} \notag \\
\delta_{n-2}\left(w_{n-3}\right) + \delta_{n-2}\left(w_{n-2}\right) &= \epsilon_{n-1}\left(w_{n-2}\right)+\epsilon_{n-1}\left(w_{n-1}\right),  \label{Eq:A3-8}  \\
\delta_{n-1}\left(w_{n-2}\right) + \delta_{n-1}\left(w_{n-1}\right) &= \epsilon_{n}\left(w_{n-1}\right)+\epsilon_{n}\left(q\right),  \label{Eq:A3-9}  \\
\delta_{m+1}\left(w_{m}\right) + \delta_{m+1}\left(w_{m+1}\right) + \delta_{m+1}\left(w_{a}\right) &= \epsilon_{m-1}^{m+1}\left(w_{m-2}\right) + \epsilon_{m-1}^{m+1}\left(w_{m-1}\right) \nonumber \\
&\quad + \epsilon_{m-1}^{m+1}\left(w_{a}\right) + t_{m}.  \label{Eq:A3-10}  
\end{align}
Here, the notation $\epsilon_{m-1}^{m}$ stands for the intermediate deflection of the shim $\left( m-1 \right)$ before the crossover at the same radius of the crossover shim $m$.
Eq. (\ref{Eq:A3-10}) is added to the original equation system to account for the additional contact force $w_{a}$.
Using the above Eq. (\ref{Eq:A3-1})-(\ref{Eq:A3-10}), the matrix $\hat{A}$ of linear coefficients is built up as in Eq. (\ref{Eq:2B-3-8}) for the non-crossover case and the algebraic system
$\hat{A}\cdot\vec{w}=\vec{b}$ with
\begin{align}
  \vec{w} &= 
  \left[ {\begin{array}{c}
    w_{1}  \\
    w_{2}  \\
    \vdots \\
    w_{n-1}  \\
    w_{a}  \\
  \end{array} } \right],  \label{Eq:A3-11}  \\
    \vec{b} &= 
  \left[ {\begin{array}{c}
    0  \\
    \vdots  \\
    0  \\
    \epsilon_{n}\left(q\right)  \\
    -t_{m}  \\
  \end{array} } \right].  \label{Eq:A3-12}  
\end{align}
is solved for all unknown forces $\vec{w}$.
The above approach can be readily generalized for multiple additional contact forces or intermediate crossovers as shown in Fig. \ref{Fig:2B-3-1}.

\end{appendices}






\vspace{0.5cm}
\backmatter

\section*{Declarations}

\bmhead{Funding}
This work was enabled by computational resources and research budget from KTM and KTM TECHNOLOGIES.

\bmhead{Acknowledgements}
Johannes Wimmer is gratefully acknowledged for providing the test bench data for validation.

\bmhead{Competing Interests}
There are no conflicting interests that would have biased this work.

\bmhead{Data Availability}
The datasets generated and analysed during the current study are not publicly available, but are available from the corresponding author on reasonable request.

\end{document}